\documentclass[preprint,12pt]{elsarticle}
\usepackage{setspace}
\usepackage{color}
\usepackage{lipsum}
\usepackage{amsfonts,amsmath,amssymb,mathrsfs}
\usepackage{epstopdf}
\usepackage{natbib}
\usepackage{algorithmic}
\usepackage{amsopn}
\usepackage{mathtools}
\usepackage{cases}
\usepackage{color}
\usepackage{enumerate}
\usepackage{graphicx}
\usepackage{subcaption}
\newtheorem{theorem}{Theorem}[section]
\newtheorem{lemma}{Lemma}[section]

\newtheorem{remark}{Remark}[section]
\newtheorem{assumption}{Assumption}

\newtheorem{proposition}{Proposition}[section]

\newtheorem{example}{Example}[section]
\usepackage{geometry}
\usepackage[none]{hyphenat}
\usepackage{titlesec}

\newcommand{\pp}{\mathbb{P}}

\renewcommand{\epsilon}{\varepsilon}

\newcommand{\R}{\mathbb{R}}
\newcommand{\Q}{\mathbb{Q}}
\newcommand{\E}{\mathbb{E}}
\newcommand{\F}{\mathcal{F}}
\newcommand{\y}{\bar{Y}}
\newcommand{\z}{\bar{Z}}
\numberwithin {equation} {section}
\DeclareMathOperator{\essinf}{ess~inf}
\journal{}

\begin{document}
	
\begin{frontmatter}
	\title{Maximum principle for robust utility optimization via Tsallis relative entropy}
	\author[math1]{Xueying Huang}
	\ead{ts23080012a31ld@cumt.edu.cn}
	\author[math2]{Peng Luo}
	\ead{peng.luo@sjtu.edu.cn}
	\author[math1]{Dejian Tian\corref{correspondingauthor}}
	\ead{djtian@cumt.edu.cn}
	\address[math1]{School of Mathematics, China University of Mining and Technology, Xuzhou, P.R. China}
	\address[math2]{School of Mathematical Sciences, Shanghai Jiao Tong University, Shanghai, 200240, P.R. China}

	\cortext[correspondingauthor]{Corresponding author}
	
	\begin{abstract}
		This paper investigates an optimal consumption-investment problem featuring recursive utility via Tsallis relative entropy. We establish a fundamental connection between this optimization problem and a quadratic backward stochastic differential equation (BSDE), demonstrating that the value function is the value process of the solution to this BSDE. Utilizing advanced BSDE techniques, we derive a novel stochastic maximum principle that provides necessary conditions for both the optimal consumption process and terminal wealth. Furthermore, we prove the existence of optimal strategy and analyze the coupled forward-backward system arising from the optimization problem.
	\end{abstract}
	
\end{frontmatter}
	
	\section{Introduction}
	The utility maximization problem constitutes a fundamental problem in financial mathematics. Merton \cite{merton1971} pioneered the study of the optimization problem with standard utility and linear wealth in complete market. Since then, generally through the Bellman approach and the martingale method, there have been many generalizations of Merton's problem. Schroder and Skiadas \cite{schroder1999} generalized the martingale method based on the continuous recursive utility from Duffie and Epstein \cite{duffie1992} to compute portfolios and consumption plans,
	and used the backward stochastic differential equation (BSDE) theory to obtain closed solutions for a class of stochastic differential utility (SDU).
	
	Thereafter, combined with the theory of BSDE, many scholars have carried out a wide range of investigations on the robust utility maximization problems.
	El Karoui, Peng, and Quenez \cite{el2001} discussed the recursive utility optimization problems with nonlinear wealth processes. In this work, they emphasized the symmetry between utility and wealth, showing that the problem can be well expressed by BSDEs. Thus, they established the relationship between the recursive utility problem and BSDEs. On this basis they derived a maximum principle which gives necessary conditions for optimality. Later, Skiadas \cite{skiadas2003} gave a dynamic form of the robust control problem by means of BSDEs in the model uncertainty context. 
	
	Bordigoni, Matoussi, and Schweizer \cite{bordigoni2007stochastic} studied the stochastic control problem in the context of utility maximization under model uncertainty. The latter is formulated as a $\sup-\inf$ problem $\sup_{\pi}\inf_{\Q} ~U(\pi, \Q)$, where $\pi$ belongs to a set of strategies, $\Q$ runs through a family of probability measure, and $U(\pi, \Q)$ is the sum of an expected utility term over $\Q$ and a penalization term based on \textit{classical relative entropy}. They proved that there exists a unique optimal probability measure $\Q^{*}$ for this problem, and used stochastic control method to study the dynamic value of the minimization problem. For the case of continuous filtration, they further showed that the value function is the first component of the unique solution to a BSDE with quadratic growth, generalizing the results of Skiadas \cite{skiadas2003}. Subsequently, based on the above model, Faidi, Matoussi, and Mnif \cite{faidi2011maximization} studied the maximum principles of the $\sup-\inf$ problem in complete market, and Matoussi, Mezghani, and Mnif \cite{matoussi2015robust} further considered a robust utility maximization problem under convex constraints. 
	
	Recently, an increasing number of \textit{general relative entropy} models have been investigated in the economic literature, including the works of Meyer and Gohde \cite{meyer2019generalized}, Ma and Tian \cite{ma2021generalized} and Tian \cite{tian2023pricing} for Tsallis relative entropy to study the pricing theory or risk measures, and Maenhout, Vedolin, and Xing \cite{maenhout2025robustness} for Cressie-Read divergence (a generalized relative entropy close to Tsallis relative entropy) to understand portfolio selection and general equilibrium asset pricing, and so on.   

	In our paper, we consider the robust utility maximization problem by incorporating the Tsallis relative entropy instead of the classical relative entropy. The specific formulation of the $\sup-\inf$ problem associated with Tsallis relative entropy is given by
	\begin{align*}
		\sup_{(c,\xi) \in \hat{\mathcal{A}}(x)}\underset{\Q \in \mathcal{Q}_{f}}{\inf}\left(\E_{\mathbb{P}}\Big[(D^{\Q}_T)^q h(\xi) +\int_{0}^{T}(D^{\Q}_{s})^q u(c_s) d s \Big]+\frac{1}{\gamma} H_q(\Q \mid \pp)\right),
	\end{align*}
	where $q>0$ and $q \neq 1$, $\gamma >0$ is the ambiguity aversion coefficient , $D^{\Q}$ is the density process defined by \eqref{D}, $\xi$ is a $\F_T$-measurable random variable, $h(\cdot)$ and $u(\cdot)$ are utility functions, and $H_q(\Q \mid \pp)$ represents the Tsallis relative entropy, characterizing the discrepancy, was introduced by Tsallis \cite{tsallis2009nonextensive} from the viewpoint of statistical physics. 
	Here, $\hat{\mathcal{A}}(x)$ is the admissible strategy set, $\mathcal{Q}_{f}$ is the set of robust uncertainty probability measures, and the parameter $q$ acts as a distortion parameter, governing a nonlinear transformation of probability measures, and adjusting the weight of deviations between the reference measure $\pp$ and alternative measures. In particular, when $q=1$, our model is reduced to the model of Faidi, Matoussi, and Mnif \cite{faidi2011maximization}, where $H_1(\Q\mid\pp)$  is exactly the formulation of Hansen and Sargent \cite{hansen2001}.
    For $q>1$, the agent assigns higher penalty weights to the discrepancies between $\Q$ and $\pp$; for $0<q<1$, it enhances sensitivity to the rare but extreme deviations.

	The main results of this paper are summarized as follows. Firstly, in complete market, we focus on a class of recursive utility problem, i.e. Inner Problem \eqref{eq inf}, where the agent optimizes over consumption and terminal wealth while accounting for model uncertainty via Tsallis relative entropy. Our work addresses great challenges posed by consumption term and unbounded terminal in the highly nonlinear setting of parameter $q$. To tackle these difficulties, we introduce a suitable transformation that reformulates the $f(y)|z|^{2}$-type BSDE to the one satisfying proper monotonicity conditions via It\^{o}'s formula, which allows us to directly establish the existence and uniqueness of its solution. In addition, the unbounded terminal condition brings additional difficulty, which we overcome through some techniques to derive the relationship between the robust problem \eqref{eq inf} and BSDE \eqref{eq3.1} with quadratic growth generator. More significantly, Theorem \ref{th1} shows that the value function in problem \eqref{eq inf} is the value process of the solution to the associated quadratic BSDE \eqref{eq3.1}, and the essential infimum of this problem can be attained.  Our results non-trivially extend the work of Faidi, Matoussi, and Mnif \cite{faidi2011maximization}  to the context of Tsallis relative entropy, and generalize the findings of Tian \cite{tian2023pricing}  by incorporating a consumption term and relaxing the boundedness constraint on terminal wealth.

	Second, adopting the idea from El Karoui, Peng, and Quenez \cite{el1997backward}, we derive a maximum principle (Theorem \ref{th2}) that characterizes the optimal strategy for the Outer Problem \eqref{eq4.1}, namely, the necessary conditions that optimal consumption $c^0$ and terminal wealth $\xi^0$ must satisfy. This task is highly nontrivial due to the stringent Lipschitz condition required in El Karoui, Peng, and Quenez \cite{el1997backward}, which does not naturally align with our setting. To overcome this, we carefully adapt our approach by extensively applying the continuous dependence of the transformed BSDE with monotonicity conditions to prove Proposition \ref{pro1}, and finally obtain Theorem \ref{th2}. To the best of our knowledge, this maximum principle presented here appears to be the first such result established in this framework.
	
	Finally, Theorem \ref{th exist} establishes the existence of an optimal strategy for problem \eqref{eq4.1}. We first transform the associated BSDE \eqref{eq4.2} and, similarly, reformulate problem \eqref{eq4.1} as the minimization problem \eqref{eq4.4}. Due to the constraint on the wealth process $X^{c,\xi}_t$ at time $t \in [0,T]$  in problem \eqref{eq4.4}, we further convert it into an unconstrained Auxiliary Problem \eqref{eq4.5}. Under an explicit compactness assumption on a regular minimizing sequence, we prove existence and uniqueness of an optimal strategy $(c^*,\xi^*)$ for this auxiliary problem, thereby obtaining the result of Theorem \ref{th exist}. Furthermore, based on the maximum principle (Theorem \ref{th2}) and existence of optimal strategy (Theorem \ref{th exist}), we derive the representation of the optimal consumption and terminal wealth in terms of the utility function, which leads to a characterization of the optimal strategy associated with a forward-backward system.
	
	The paper is organized as follows. Section 2 describes the model and the optimization problem. Section 3 is devoted to establish the relationship between the robust utility problem and a quadratic BSDE. In section 4, we derive the maximum principle. Finally, we investigate the existence of optimal strategy and relate the optimal strategy to the solution of a forward–backward system. Some proofs are relegated to Appendix.

	\section{The problem formulation}
	\subsection{Model uncertainty setup and Tsallis relative entropy}
	Let $T\in(0,+\infty)$ be a finite time horizon and  $(\Omega,\mathcal{F},(\mathcal{F}_t)_{0\leq t\leq T}, \mathbb{P})$ be a complete filtered probability space. $ (\mathcal{F}_t)_{0\leq t\leq T}$ is the completed natural $\sigma$-algebra filtration generated by $d$-dimensional Brownian motion $(B_t)_{0\leq t\leq T}$ , which is continuous. 
	
	For any two probability measures $\Q \ll \mathbb{P}$ on $\F_T$, the density process of $\Q$ with respect to $\mathbb{P}$ is the continuous martingale $D^{\Q}$ as follows:
	\begin{align}\label{D}
		D_t^{\Q}=\E_{\mathbb{P}} \left[\frac{d\Q}{d\mathbb{P}} \mid \F_t \right], ~~~t\in [0,T].
	\end{align}  
	For any probability measure $\Q$ equivalent with respect to $\mathbb{P}$, there exists a unique progressively measurable $\R^d$-valued process $(\eta_{t})_{t\in [0,T]}$ such that $\int_{0}^{\cdot}\eta_sdB_s$ is a local martingale on $[0,T]$ under $\mathbb{P}$ and 
	\begin{equation}\label{eq:unique-eta}
		D_t^{\Q}=\mathcal{E}(\eta \cdot B)_t, ~~~ t\in [0,T],
	\end{equation}
	where $\mathcal{E}$ denotes the stochastic exponential.  We denote by $D^{\Q}_{t,s}=D^{\Q}_s/D^{\Q}_t$, $0 \leq t \leq s \leq T$. At $t=0$, the notation $D^{\Q}_{0,s}$ will be simplified by $D^{\Q}_{s}$, $s \in [0,T]$.

From Tsallis \cite{tsallis2009nonextensive}, we know that Tsallis relative entropy can characterize the distance for two probability measures. Specifically,  for any two probability measures $\Q$ and $\mathbb{P}$ on $\left(\Omega, \mathcal{F}_{T}\right)$, the $q$-generalization of the relative entropy, i.e. Tsallis relative entropy, is defined as follows
	$$H_{q}(\Q \mid \mathbb{P}):=\left\{\begin{array}{ll}
		\E_{\mathbb{P}}\left[\left(D_{T}^{\Q}\right)^{q} \ln _{q}\left(D_{T}^{\Q}\right)\right], & \Q \ll \mathbb{P}, \\
		+\infty, & others,
	\end{array}\right.$$
	where $q>0$  and  $q \neq 1$,  $\ln_{q}$  is  $q$-logarithm, which is defined by
	$$\ln _{q}(x):=\left\{\begin{array}{ll}
		\frac{x^{1-q}-1}{1-q}, & x \geq 0 \quad and \quad 0<q<1, \\
		\frac{x^{1-q}-1}{1-q}, & x>0 \quad and \quad q>1 .
	\end{array}\right.$$
	The inverse function, i.e.  $q$-exponential is
	$$\exp _{q}(x):=\left\{\begin{array}{ll}
		{[1+(1-q) x]^{\frac{1}{1-q}},} & x\geq-\frac{1}{1-q} \quad and \quad 0<q<1, \\
		{[1+(1-q) x]^{\frac{1}{1-q}},} & x<-\frac{1}{1-q} \quad and \quad q>1 .
	\end{array}\right.$$
Similarly, if $\Q \ll \mathbb{P}$ on $\mathcal{F}_T$,  we can define the conditional Tsallis relative entropy between $\Q$ and $\mathbb{P}$ under $\mathcal{F}_t$ by
	$$H_{q, t}(\Q \mid \mathbb{P}):=\E_{\mathbb{P}}\left[\left.\left( D_{t,T}^{\Q}\right)^{q} \ln _{q} D_{t,T}^{\Q} \right\rvert\, \mathcal{F}_{t}\right],~~~t\in[0,T].$$ 	
	\begin{remark}
		For any  $q>0$ and $q \neq 1$, $\ln_{q}(\cdot)$  and  $\exp _{q}(\cdot)$ are well-defined. It can be easily verified that $\ln_{q}(x) \to \ln(x)$, and $\exp_{q}(x) \to \exp(x)$, as $q$ goes to 1.
	\end{remark}

The uncertainty probability measure set is defined by 
		\begin{align}\label{eq:qf}
        \mathcal{Q}_f:=\left\{\Q  \;\middle|\;    
		\Q\sim\pp ~ \textrm{on} ~ \mathcal{F}_{T}, ~ H_{q}(\Q \mid \pp)<+\infty 
		\right\}.\end{align}
Uncertainty sets of this type are increasingly being employed to address various issues in mathematical finance and economics field. For example, Meyer-Gohde \cite{meyer2019generalized} utilizes them to match detection‑error probabilities, thereby yielding comparable asset‑pricing implications across models; Tian \cite{tian2023pricing} applies them to the pricing of contingent claims; and Maenhout, Vedolin, and Xing \cite{maenhout2025robustness} adopt them to study portfolio choice and general equilibrium asset pricing, using the Cressie–Read divergence—a generalized relative entropy that is closely related to Tsallis relative entropy; and Bo et al. \cite{bo2026} employ them for reinforcement learning in jump‑diffusion models.

\subsection{The financial market and related optimal problem}
For $\beta>1$, let $L^{\beta}_+(\F_T)$ be the set of all nonnegative $\F_T$-measurable random variable, satisfying $\E_\pp[\xi^{\beta}]<+\infty$.
	$\mathcal{C}^{\beta}$ represents the set of all nonnegative progressively measurable process $(c_t)_{0\leq t\leq T}$, satisfying $\E_\pp[\int_{0}^{T}(c_t)^{\beta}dt]<+\infty$.  $\mathcal{S}^\beta$ is the set of all $(\F_t)$-adapted and continuous process $(Y_t)_{0\leq t\leq T}$ satisfying 
	$ \E_\pp\left[\sup_{0\leq t\leq T}|Y_t|^{\beta}\right]<+\infty$, and $\mathcal{M}^{\beta}$ represents the set of all $(\F_t)$-progressively measurable $\R^{d}$-valued process $(Z_t)_{0\leq t\leq T}$ such that
	$\E_\pp[(\int_0^T|Z_s|^2ds)^\frac{\beta}{2}]<+\infty$.

	Considering a complete market consists of a risk-free bond with  interest rate process $r=0$ and $d$ stocks, and the price process for the $i$th stock is modeled by the stochastic differential equation
	$$d S_{t}^{i}=S_{t}^{i}\left(b_{t}^{i} d t+\sigma_{t}^{i} d B_{t}\right), \quad i=1, \cdots, d,$$
	where $b^i$ is a $\R$-valued bounded adapted process, standing for the average return of the $i$th risky asset at time $t$, and the progressively measurable process $\sigma^{i}$ taking values in $\R^{1 \times d}$ denotes the  volatility in the $i$th risky assets at time $t$. The volatility matrix is denoted by $\sigma=   \left(\left(\sigma^{1}\right)^{*}, \cdots\left(\sigma^{d}\right)^{*}\right)^{*}$ with full rank, which is invertible and bounded. Let $\theta=  \sigma^{-1} b$ be an $\R^{d}$-valued, progressively measurable and bounded process representing the risk premium of the financial market.  Since the market is complete, there exists a unique equivalent local martingale measure $\tilde{\pp}$, whose density process is given by the stochastic exponential 
	$\mathcal{E}(-\theta \cdot B)$.

	Let $\pi^i_t$ be the amount invested in the $i$th risky asset at time $t$. Suppose that the investment strategy $\pi_{t}=   \left(\pi_{t}^{1}, \cdots, \pi_{t}^{d}\right), t \in[0, T]$ is in the set $\Pi$,  which represents the space of all progressively measurable process $\pi=(\pi_t)_{0 \leq t \leq T}$ satisfying $\int_{0}^{T}|\pi_t \sigma_t|^2dt < +\infty$.  We also assume that the consumption rate is $c=(c_t)_{0\leq t\leq T}$. Given initial wealth $x>0$, then, the wealth process is $(X_{t})_{t \in [0,T]}$ satisfying the following equation
	$$ dX^{x,c,\pi}_t= \pi_{t}\sigma_{t}(\theta_t dt+dB_t)-c_tdt, ~~X^{x,c,\pi}_0=x.$$

	From  El Karoui, Peng and Quenez \cite{el2001}, rather than treating the portfolio process as control, one may alternatively formulate the  problem in terms of terminal wealth $\xi$. Thus, the wealth process $X^{c,\xi}_t$ associated with a given strategy $(c,\xi)$ is the solution to the BSDE below:
	\begin{equation}\label{eq wealth}
		\left\{\begin{array}{ll}
			d X_{t}^{c, \xi} &=(\pi_{t}\sigma_{t}\theta_t-c_t)dt+\pi_{t}\sigma_{t}dB_t, \quad t \in[0, T], \\
			X_{T}^{c, \xi}&=\xi.
		\end{array}\right.
	\end{equation}
For $(c,\xi)\in \mathcal{C}^\beta\times L^\beta_+(\mathcal F_T)$, $\beta>1$, equation \eqref{eq wealth} determines a unique wealth process and replicating exposure
$$(X^{c,\xi},\pi^{c,\xi}\sigma)\in\mathcal{S}^\beta\times \mathcal{M}^\beta.$$
Then it can uniquely determine the portfolio strategy $\pi^{c,\xi}$. By martingale argument, it is easy to verify that 
$$
X_t^{c,\xi}=\mathbb E_{\widetilde{\mathbb P}}\left[\xi+\int_t^T c_sds\mid\mathcal F_t\right]\ge0 \text{~~~and~~~}
X_0^{c,\xi}=\mathbb E_\pp\left[\widetilde D_T\xi+\int_0^T\widetilde D_t c_tdt\right].
$$
Therefore, $(c,\xi)$ provides a terminal-wealth formulation for
consumption--portfolio strategies $(c,\pi)$. Appendix A establishes the equivalence of the
associated optimization problems in suitable admissible class.

The following assumptions are imposed on the utility functions $u(\cdot)$ and $h(\cdot)$.
	\begin{assumption}\label{a2}
		The functions $u,h:[0,+\infty)\to[0,+\infty)$ are finite and continuous, strictly increasing and strictly concave, and $C^2$ on $(0,+\infty)$.
	\end{assumption}
	\begin{assumption}\label{a3}
		 The utilities $u(\cdot)$ and $h(\cdot)$  satisfy the Inada conditions, i.e. $$u^{\prime}(+\infty)=h^{\prime}(+\infty)=0,~~ u^{\prime}(0+)=h^{\prime}(0+)=+\infty.$$
	\end{assumption}
	\begin{assumption}\label{a3-1}
There exist $a\in(0,1)$ and $K<\infty$ such that
$u(z)+h(z)\le K(1+z^a)$,$z\ge0$.
\end{assumption}

We will give the definition of admissible strategy set below.  Under the above assumptions, fix $q>1$, $\gamma>0$, $p>q+1$, and $\beta>\max\{1,2ap\}$. Consequently,  for any $(c,\xi)\in \mathcal{C}^\beta\times L^\beta_+(\mathcal F_T)$, then $\mathbb{E}_\pp[\int_0^Tu(c_s)^{2p}ds+h(\xi)^{2p}]<+\infty$. Define
\begin{align*}
   \mathcal{A}:=&\left\{(c,\xi)\in \mathcal{C}^\beta\times L^\beta_+(\mathcal F_T):
\mathbb E_\pp\big[\int_0^T u'(c_t)^{2p}dt+ h'(\xi)^{2p}\big]<+\infty\right\}, \\
\mathcal{\hat{A}}(x):=&\{(c,\xi)\in \mathcal{A}:X_0^{c,\xi}\le x\},\qquad x>0.
\end{align*}
Here $X_0^{c,\xi}$ is the initial value of \eqref{eq wealth}, with the state-price representation above. 

\begin{remark}
The sets $\mathcal{A}$ and $\hat{\mathcal{A}}(x)$ are convex. Indeed, for $c^\alpha=(1-\alpha)c^0+\alpha c^1$ between two admissible endpoints, one has
$$
u'(c^\alpha)^{2p}\le u'(c^0)^{2p}+u'(c^1)^{2p} \text{~~~and ~~~}
u(c^\alpha)^{2p}\le u(c^0)^{2p}+u(c^1)^{2p},
$$
and the analogous bounds also hold for $h$ and $h'$. These inequalities follow from the monotonicity of the utilities and
their derivatives.  The linearity of wealth budget equation implies the convexity of $\hat{\mathcal{A}}(x)$. The same bounds imply the uniform integrability
of the $p$th powers of the utilities and their derivatives along every
fixed segment in $\mathcal{A}$. Obviously,  $\hat{\mathcal{A}}(x)$ is nonempty for every $x>0$, for example $(c,\xi)=(\frac{x}{2}, \frac{x}{2T})\in\hat{\mathcal{A}}(x)$.

Assumption A3 controls the growth rate of the utilities, whereas $\mathcal{A}$ imposes the required marginal-utility moments. The uniform moment bound and stability under limits used in the
existence proof are provided by assumption A4 in Section \ref{sec:5}.
\end{remark}

We now can formulate and analyze the following robust optimization problem involving both consumption processes $(c_t)_{t\in [0,T]}$ and terminal wealth $\xi$ via Tsallis relative entropy.  More precisely, for any initial wealth $x>0$, 
	we will focus on the $\sup-\inf$ optimal problem as follows
	\begin{align}\label{eq:main-problem-static}
		\hat{V}(x)=\sup_{(c,\xi) \in \hat{\mathcal{A}}(x)} \underset{\Q \in \mathcal{Q}_{f}}{\inf}\left(\E_{\mathbb{P}}\left[\left(D_{ T}^{\Q}\right)^{q}h(\xi)+\int_{0}^{T}\left(D_{ s}^{\Q}\right)^{q} u\left(c_{s}\right) d s \right]+\frac{1}{\gamma} H_{q}(\Q \mid \pp)\right),
	\end{align}
	where $\gamma>0$ is the ambiguity aversion coefficient, $u(\cdot)$ and $h(\cdot)$ are utility functions for consumption rate process and terminal wealth  respectively,  and $\hat{\mathcal{A}}(x)$ is the set of admissible strategies and $\mathcal{Q}_f$ is the collection of uncertainty probability measures given by \eqref{eq:qf}.
	Our primary objective is to  characterize the optimal consumption and terminal wealth.
	
	\begin{remark}
		For the case $q=1$, the problem \eqref{eq:main-problem-static} reduces to the
		classical robust control framework investigated by Bordigoni, Matoussi and Schweizer \cite{bordigoni2007stochastic} or Faidi, Matoussi and Mnif \cite{faidi2011maximization}, where $H_1(\Q \mid \pp)$ is the relative entropy. The parameter $q$ can be viewed as a bias or distortion of the original probability measure, the agent  puts less/more weight on events with large $D_{\cdot}^{\Q}$ for different $q$. For more economical explanations, one can refer to  Meyer-Gohd \cite{meyer2019generalized}.  We mainly consider $q>1$ throughout the paper, see Remark \ref{rem:qless1} for the discussion on $0<q<1$.  

	\end{remark}
	
	\begin{remark}
		When the ambiguity aversion coefficient $\gamma$ goes to zero, the agent completely trusts the reference measure $\pp$, reducing it to the classical expected utility maximization problem; while as $\gamma$ goes to $+\infty$, the entropy penalty term becomes negligible, and the problem prioritizes protection against the worst-case scenario across of all uncertainty probability measures, leading to extreme conservatism. 
	\end{remark}

	\section{The robust utility problem via Tsallis relative entropy and corresponding BSDE}
	
	In this section, we will focus on the inner infimal problem. Our objective is to establish the relationship between the robust utility problem and a special kind of quadratic BSDE. To achieve this,  we consider the following dynamic version of the inner problem as follows: for any $t\in[0,T]$, 
	\begin{align}\label{eq inf}
		\textbf{Inner Problem}:
		\underset{\Q \in \mathcal{Q}_{f}}{\essinf}\left(\E_{\mathbb{P}}\left[\left(D_{t, T}^{\Q}\right)^{q} \zeta +\int_{t}^{T}\left(D_{t, s}^{\Q}\right)^{q} U_s d s \mid \mathcal{F}_{t}\right]+\frac{1}{\gamma} H_{q, t}(\Q \mid \pp)\right),
	\end{align}
	where $\zeta$ and $U$ are suitable nonnegative random variable and process respectively.  
	
	When $q=1$, the above inner  problem is studied by  Bordigoni, Matoussi, and Schweizer \cite{bordigoni2007stochastic} and Faidi, Matoussi, and Mnif \cite{faidi2011maximization},  they investigated a recursive utility problem within the framework of classical relative entropy, where the associated BSDE has a standard quadratic growth in $z$. We will see that the structure of our problem and its corresponding quadratic BSDE are significantly different from theirs.  
	
\begin{theorem}\label{th1}
Let $\gamma>0$, and $q>1$. Suppose that a nonnegative random variable $\zeta$ and a nonnegative adapted process $U$ satisfy $\E_{\pp}[\zeta+\int_0^TU_sds]<+\infty$.
Then we have  
\begin{align}\label{eq:bianhuaqian}
Y_t=\underset{\Q \in \mathcal{Q}_{f}}{\essinf}\left(\E_{\mathbb{P}}\left[\left(D_{t, T}^{\Q}\right)^{q} \zeta +\int_{t}^{T}\left(D_{t, s}^{\Q}\right)^{q}  U_s d s \mid \mathcal{F}_{t}\right]+\frac{1}{\gamma} H_{q, t}(\Q \mid \pp)\right), \quad t\in [0,T],  
\end{align}
		where $(Y,Z)$  is the  unique solution of the  following BSDE 
		\begin{equation}\label{eq3.1}
			Y_t=\zeta-\int_t^T\left(\frac{\gamma}{2} \frac{\left|Z_{s}\right|^{2}}{\mu\left(Y_{s}\right)}-U_s \right)ds-\int_t^TZ_{s} d B_{s},~~\quad t \in[0, T],
		\end{equation}
		such that $Y$ is continuous,  $\exp _{q}(-\gamma Y)$ is of class $D$, and $\int_0^TZ_s^2ds<+\infty$, where
		\begin{align*}
			\mu(x):=\frac{1}{q} \left(1-(1-q)\gamma x \right)
			=\frac{1}{q} \left(\exp_{q}(-\gamma x) \right)^{1-q}.
		\end{align*}
		Moreover, the essential infimum of problem \eqref{eq:bianhuaqian} can be attained.
	\end{theorem}
	
\noindent{\textbf{Proof.}}	
		We will divide the proof into the following three steps.
		
		\textbf{Step 1:} We claim that BSDE \eqref{eq3.1} has a unique solution $(Y,Z)$ such that $Y$ is continuous,  $\exp _{q}(-\gamma Y)$ is of class $D$, and $\int_0^TZ_s^2ds<+\infty$. 
		
		Define $\check{Y}:=\exp _{q}(-\gamma Y)$, where $Y$ satisfies BSDE $\eqref{eq3.1}$. Applying It\^{o}'s formula to $\check{Y}$, we have
		\begin{align}\label{eq change}
			\check{Y}_{t}=\exp_{q}(-\gamma \zeta)-\int_t^T \gamma \check{Y}_{s}^{q} U_s d s-\int_t^T\check{Z}_s d B_{s}, \quad t \in[0, T],
		\end{align}
		where $\check{Z}=-\gamma \check{Y}^{q} Z$. Due to the facts that  $q>1$ and $U\geq0$, then the generator of BSDE \eqref{eq change} satisfies the monotonicity condition. 
		Since $\zeta\geq0$ and $q>1$, it implies $\exp_{q}(-\gamma \zeta)\in(0,1)$, combining $\mathbb{E}[\int_0^TU_sds]<+\infty$, 
		then there exists a unique $L^1$ solution $(\check{Y}, \check{Z})$ to BSDE \eqref{eq change}, in which $\check{Y}$ is continuous, strictly positive, of class $D$,   and $\int_0^T\check{Z}_s^2ds<+\infty$
		according to Proposition 2.2 in Xing \cite{xing2017}. Moreover, it also implies that $\mathbb{E}_{\pp}[\int_0^T\check{Y}_{s}^{q} U_sds]<+\infty$.  Therefore, the claim is proved.

		\textbf{Step 2:} For any $t\in[0,T]$, we show 
		\begin{align}\label{eq:xiaoyudengyu}
			Y_{t} \leq \underset{\Q \in \mathcal{Q}_{f}}{\essinf}\left(\E_{\mathbb{P}}\left[\left(D_{t, T}^{\Q}\right)^{q} \zeta +\int_{t}^{T}\left(D_{t, s}^{\Q}\right)^{q}  U_s d s \mid \mathcal{F}_{t}\right]+\frac{1}{\gamma} H_{q, t}(\Q \mid \pp)\right).
		\end{align}
        
		Case (i): we assume that $\zeta$ and $U$ is bounded above. Obviously, in this case, $Y$ is also bounded. 
        
		For any given $\mathbb{Q}\in\mathcal{Q}_f$, since $H_q(\mathbb{Q}|\mathbb{P})<+\infty$,  we have $(D_T^\mathbb{Q})^q\in L^1$. For $q>1$, by Jensen's inequality we know that the process $(D^\mathbb{Q})^q$ is a uniformly integrable submartingale under $\pp$.  Besides,   by \eqref{eq:unique-eta}, there exists a unique $\eta$ such that $$ dD_{s}^\mathbb{Q}=D_{s}^\mathbb{Q} \eta_{s} dB_{s}, \quad D_{0}^\mathbb{Q}=1.$$
		Applying It\^{o}'s formula to $(D^\mathbb{Q})^{q}$, it implies
		\begin{align*}
			d(D_{s}^\mathbb{Q})^{q} &=q(D_{s}^\mathbb{Q})^{q-1}dD_{s}^\mathbb{Q}+\frac{1}{2}q(q-1)(D_{s}^\mathbb{Q})^{q-2}\left(dD_{s}^\mathbb{Q}\right)^{2} \\
			&=(D_{s}^\mathbb{Q})^{q}\left(\frac{1}{2}q(q-1) \eta_{s}^{2} d s+q \eta_{s} dB_{s}\right).
		\end{align*}Furthermore, 	using It\^{o}'s formula to $Y(D^\mathbb{Q})^{q}$, we derive that
		\begin{align*}
			dY_{s} (D_{s}^\mathbb{Q})^{q} =& Y_{s} d(D_{s}^\mathbb{Q})^{q}+(D_{s}^\mathbb{Q})^{q} dY_{s}+d(D_{s}^\mathbb{Q})^{q} dY_{s} \\
			=&Y_{s}(D_{s}^\mathbb{Q})^{q}\left(\frac{1}{2} q(q-1) \eta_{s}^{2} ds+q\eta_{s} dB_{s}\right) \\
			&+(D_{s}^\mathbb{Q})^{q}\left(\frac{\gamma}{2} \frac{\left|Z_{s}\right|^{2}}{\mu\left(Y_{s}\right)} ds - U_s ds+Z_{s} dB_{s}\right)+q(D_{s}^\mathbb{Q})^{q} \eta_{s} Z_{s}ds\\
			=&(D_{s}^\mathbb{Q})^{q}\Gamma_s ds +dL_s,
		\end{align*}
		where $L_{s}=\int_{0}^{s} (D_{v}^\mathbb{Q})^{q}\left(qY_{v} \eta_{v}+Z_{v}\right) dB_{v} $ and  
		$$\Gamma_{s}:=\frac{1}{2} q(q-1) Y_{s} \eta_{s}^{2}+\frac{\gamma}{2}\frac{\left|Z_{s}\right|^{2}}{\mu\left(Y_{s}\right)}-U_s+q \eta_{s} Z_{s}.$$
		
		Since $Y_{t}=\frac{q  \mu\left(Y_{t}\right)-1}{\gamma(q-1)}$, by direct calculations, one can get, 
		\begin{align*}
			\Gamma_{t}
			=\frac{ \mu\left(Y_{t}\right)}{2\gamma}\left(q \eta_{t}+\frac{\gamma Z_{t}}{ \mu\left(Y_{t}\right)}\right)^{2}-\frac{q}{2 \gamma} \eta_{t}^{2}-U_t.
		\end{align*}
		Therefore,  
		\begin{align}\label{eq:optimal-q}
			d Y_{t} (D_{t}^\mathbb{Q})^{q} 
			=(D_{t}^\mathbb{Q})^{q}\left[\frac{ \mu\left(Y_{t}\right)}{2 \gamma}\left(q \eta_{t}+\frac{\gamma Z_{t}}{ \mu\left(Y_{t}\right)}\right)^{2}-\frac{q}{2 \gamma} \eta_{t}^{2}-U_t \right] d t+dL_{t}.
		\end{align}
		
		For $t<T$, let $\tau_{n}\geq t$ be a sequence of stopping times converging to $T$ such that $L_{\tau_{n}}-L_t$ is a true martingale. Then we derive that
		\begin{align}\label{eq3.2}
			Y_{t} =&\mathbb{E}_{\pp} \left[Y_{\tau_{n}} (D_{t, \tau_{n}}^\mathbb{Q})^{q} \mid \mathcal{F}_{t}\right]+\frac{q }{2 \gamma} \mathbb{E}_{\pp} \left[\int_{t}^{\tau_{n}} \eta_{s}^{2} (D_{t, s}^\mathbb{Q})^{q} ds \mid \mathcal{F}_{t}\right] \nonumber \\
			&-\frac{1}{2 \gamma} \mathbb{E}_{\pp} \left[\left.\int_{t}^{\tau_{n}} \mu\left(Y_{s}\right) (D_{t, s}^\mathbb{Q})^{q}\left(q \eta_{s}+\frac{ \gamma Z_{s}}{ \mu\left(Y_{s}\right)}\right)^{2} d s \right\rvert\, \mathcal{F}_{t}\right] 
			+\mathbb{E}_{\pp} \left[\int_{t}^{\tau_{n}} (D_{t, s}^\mathbb{Q})^{q} U_s ds \mid \mathcal{F}_{t}\right].
		\end{align}
Since $Y$ and $U$ are bounded, one can get that
$(D^{\Q})^q Y+\int_{0}^{\cdot}(D^{\Q}_s)^q U_sds$ is of class $D$. By dominated convergence theorem and monotone convergence theorem, when $n \to \infty$, we have
		\begin{align*}
			Y_{t} = & \mathbb{E}_{\pp} \left[Y_{T} (D_{t, T}^\mathbb{Q})^{q} \mid \mathcal{F}_{t}\right]+\mathbb{E}_{\pp} \left[\int_{t}^{T} (D_{t, s}^\mathbb{Q})^{q} U_s ds \mid \mathcal{F}_{t}\right]+\frac{q}{2\gamma} \mathbb{E}_{\pp} \left[\int_{t}^{T} \eta_{s}^{2} (D_{t, s}^\mathbb{Q})^{q} ds \mid \mathcal{F}_{t}\right]\\
			&-\frac{1}{2 \gamma} \mathbb{E}_{\pp} \left[\left.\int_{t}^{T} \mu\left(Y_{s}\right) (D_{t, s}^\mathbb{Q})^{q}\left(q \eta_{s}+\frac{\gamma Z_{s}}{\mu\left(Y_{s}\right)}\right)^{2} d s \right\rvert\, \mathcal{F}_{t}\right].
		\end{align*}
		By Corollary 2.2 in Tian \cite{tian2023pricing}, one has 
		$H_{q, t}(\Q \mid \pp)=\frac{q}{2} \mathbb{E}_{\pp} \left[\int_{t}^{T} \eta_{s}^{2} (D_{t, s}^\mathbb{Q})^{q} ds \mid \mathcal{F}_{t}\right]$. Dropping the last negative term, taking the essential infimum over $\mathcal{Q}_f$,  
		thus, \eqref{eq:xiaoyudengyu} holds.
		
Case (ii): $\zeta$ and $U$ are not bounded.  In this case, consider $\zeta^N=\zeta\wedge N$ and $U^N=U\wedge N$ for $N\geq1$. Its corresponding solution to BSDEs \eqref{eq change} and \eqref{eq3.1} are  denoted by $(\check Y^N, \check Z^N)$ and $(Y^N, Z^N)$, respectively. By comparison theorem and stability property for BSDE \eqref{eq change} in $L^\beta (\beta>1)$ space, we know that $\check Y^N$ is decreasing in $N$ and $\check Y^N \downarrow \check Y$, and hence $Y^N\uparrow Y$. 

For the fixed $N$, we can get for all $t\in[0,T]$,
\begin{align}\label{eq:xiaoyudengyu-jieduan}
			Y^N_{t} \leq \underset{\Q \in \mathcal{Q}_{f}}{\essinf}\left(\E_{\mathbb{P}}\left[\left(D_{t, T}^{\Q}\right)^{q} \zeta^N +\int_{t}^{T}\left(D_{t, s}^{\Q}\right)^{q}  U_s^N d s \mid \mathcal{F}_{t}\right]+\frac{1}{\gamma} H_{q, t}(\Q \mid \pp)\right)
\end{align}by the result of case (i). Then one can get 
\eqref{eq:xiaoyudengyu} by letting $N$ goes to infinity. 
        
\textbf{Step 3:} We show the essential infimum in \eqref{eq:xiaoyudengyu} is actually attained. Specifically, we will show the essential infimum is attainable at  $\mathbb{Q}^{*}\in\mathcal{Q}_{f}$, where
		$$ \eta^{\mathbb{Q}^{*}}:=-\frac{\gamma Z}{q \mu(Y)}.$$

		Since $\check{Z}=-\gamma \check{Y}^{q} Z$, we know that
		\begin{align*}
			d \check{Y}_{t}=\check{Y}_{t}(\gamma \check{Y}_{t}^{q-1} U_t dt-\gamma \check{Y}_{t}^{q-1} Z_td B_{t}).
		\end{align*}Then one can get the explicitly expression 
		$$\check{Y}_{t}=\check{Y}_{0}\exp \left(\int_{0}^{t}\gamma \check{Y}_{s}^{q-1} U_s ds \right)\mathcal{E} \left(\frac{\check{Z}}{\check{Y}} \cdot B \right)_t, \quad t\in[0,T].$$Since $\check{Y}$ is of class $D$ and strictly positive,  and $U\geq0$, thus   $\mathcal{E} \left(\frac{\check{Z}}{\check{Y}} \cdot B \right)$ is a uniformly integrable martingale.  
		Moreover, we have 
		$$D_T^{\mathbb{Q}^{*}}=\mathcal{E} \left(\frac{\check{Z}}{\check{Y}} \cdot B \right)_T=\mathcal{E} \left(-\frac{\gamma Z}{q \mu(Y)} \cdot B \right)_T=\mathcal{E} \left(\eta^{\mathbb{Q}^{*}} \cdot B \right)_T.$$
		
		On the other hand,  since $\check{Y}_T\in(0,1)$, we can obtain that
		\begin{align}\label{eq:dqt}
(D_t^{\mathbb{Q^*}})^q
			=\left(\frac{\check{Y}_{t}}{\check{Y}_{0}} \right)^{q}\exp \left(-\int_{0}^{t}\gamma\frac{U_s}{\mu(Y_s)}ds \right) 
			\leq \left(\frac{\check{Y}_{t}}{\check{Y}_{0}} \right)^{q}\leq \check{Y}_{0}^{-q},
		\end{align}and when  $t=T$, then the Tsallis relative entropy $H_q(\mathbb{Q}^*|\mathbb{P})<+\infty$ since $\check{Y}_T\in(0,1)$. 
        
		Furthermore, one can also verify that $(D^{\Q^*})^q Y+\int_{0}^{\cdot}(D^{\Q^*}_s)^q U_sds$ is uniformly integrable. In fact, 
		since $(D^{\Q^*})^q$ is a uniformly integrable submartingale, it means it is also bounded by $\check{Y}_{0}^{-q}$.  Taking $\Q^*$ and coming back to the equation  \eqref{eq:optimal-q}, 
		we define the process $V^{\Q^*}$ as follows
		$$V^{\Q^*}_t=Y_{t} (D_{t}^{\mathbb{Q}^*})^{q}+\int_{0}^{t}(D_{s}^{\mathbb{Q}^*})^{q} \left[ \frac{q}{2 \gamma} \eta_{s}^{2}+U_s \right]ds, ~~~t \in [0,T]. $$
		Then we have 
		$$V^{\Q^*}_t=V^{\Q^*}_T-\int_{t}^{T}dL_s, ~~~t \in [0,T],$$
		where 
		$V^{\Q^*}_T=\zeta (D_{T}^{\mathbb{Q}^*})^{q}+\int_{0}^{T}(D_{s}^{\mathbb{Q}^*})^{q} \left[ \frac{q}{2 \gamma} \eta_{s}^{2}+U_s \right]ds.$ From 
		\eqref{eq:dqt}, one can derive that 
		$$(D_{t}^{\mathbb{Q}^*})^{q}U_t \leq \check{Y}_0^{-q} (\check{Y}_t)^q U_t, ~~t \in[0,T].$$  
		By step 1, since $\mathbb{E}_{\pp}[\int_0^T\check{Y}_{s}^{q} U_sds]<+\infty$, then $\mathbb{E}_{\pp}[\int_0^T(D_{s}^{\mathbb{Q}^*})^{q}U_sds]<+\infty$. 
		Moreover, combining the facts that $H_{q}(\Q^*|\pp)=\frac{q}{2}\E_{\pp} \left[\int_{0}^{T}(D_{s}^{\mathbb{Q}^*})^{q} \eta_{s}^{2} ds \right] < +\infty$, $(D_T^{\Q^*})^q$ is bounded and $\mathbb{E}_{\pp}[\zeta]<+\infty$, finally we obtain that $\mathbb{E}_{\pp}[V_T^{\Q^*}]<\infty$. It implies $V^{\Q^*}_t=\mathbb{E}_{\pp}[V^{\Q^*}_T|\mathcal{F}_t]$, $t\in[0,T]$ is uniformly integrable.   Hence, $(D^{\Q^*})^q Y+\int_{0}^{\cdot}(D^{\Q^*}_s)^q U_sds$ is uniformly integrable. Therefore, the essential infimum in \eqref{eq:xiaoyudengyu} is actually attained by letting $n$ go to infinity in \eqref{eq3.2}. \hfill$\Box$		
		

	By discarding the investment strategy, Theorem \ref{th1} establishes a general connection between the inner problem and  a specific BSDE.  This essential relationship provides an effective tool for our original problem. 
	
	\begin{remark}
		One can see that BSDE \eqref{eq3.1} is very different from the standard quadratic BSDE established by Bordigoni, Matoussi, and Schweizer \cite{bordigoni2007stochastic}. Theorem \ref{th1} extends their findings from the classical relative entropy framework to the Tsallis entropy setting. In contrast to Theorem 3.1 in Tian \cite{tian2023pricing}, our formulation incorporates the consumption utility process $(U_t)_{t \in [0,T]}$ and unbounded terminal value $\zeta$, leading to a more general optimization problem and broadening the applicability of the results.  See Bahlali and Tangpi \cite{bahlali2018bsdes}, Zheng, Zhang, and Feng \cite{Zheng2021}, and Imkeller, Likibi Pellat, and Menoukeu-Pamen \cite{imkeller2024differentiability} for more details of this kind of BSDEs.
	\end{remark}
	
	\begin{remark}\label{rem:qless1}

    When $q>1$, the generator of BSDE \eqref{eq change} satisfies a monotonicity condition, which readily ensures the existence and uniqueness of its solution. However, for $0<q<1$, the monotonicity argument used in this paper does not yield uniqueness, we will leave this situation as it is and study it further in the future. 
	\end{remark}
	
\section{Maximum principle for the robust utility problem via Tsallis relative entropy}\label{sec:4}
	
    
    In this section, we will derive the necessary conditions characterizing the optimal policy $(c^0, \xi^0)$, namely the maximum principle.

	Relying on Theorem \ref{th1},  we will focus on the following outer robust utility problem.
	\begin{equation}\label{eq4.1}
		\noindent \textbf{Outer Problem:} ~~~~~~
		V(x)=\sup _{(c,\xi) \in \hat{\mathcal{A}}(x)} Y_{0}^{c,\xi}, \quad x>0,
	\end{equation}
	where,  for all $t\in[0,T]$,  the dynamic
	\begin{align}\label{eq essinf}
		Y^{c, \xi}_t=\underset{\Q \in \mathcal{Q}_{f}}{\essinf}\left(\E_{\mathbb{P}}\left[\left(D_{t, T}^{\Q}\right)^{q} h(\xi) +\int_{t}^{T}\left(D_{t, s}^{\Q}\right)^{q}  u(c_s) d s \mid \mathcal{F}_{t}\right]+\frac{1}{\gamma} H_{q, t}(\Q \mid \pp)\right)
	\end{align}
	is the first component of solution of the following BSDE 
	\begin{equation}\label{eq4.2}
		\left\{\begin{array}{ll}
			d Y_{t}^{c, \xi} &=\left[\frac{\gamma}{2} \frac{\left|Z_{t}^{c, \xi}\right|^{2}}{\mu\left(Y_{t}^{c, \xi}\right)}- u(c_t) \right]d t+Z_{t}^{c, \xi} d B_{t}, \quad t \in[0, T], \\
			Y_{T}^{c, \xi}&=h(\xi),
		\end{array}\right.
	\end{equation}
	where $\xi$ is coming from \eqref{eq wealth}.
	
	Let $\bar{Y}^{c,\xi}=\exp_{q}(-\gamma Y^{c, \xi})$, $\z^{c,\xi}=-\gamma (\y^{c,\xi})^q Z^{c,\xi}$, and $g(\xi)=\exp_{q}(-\gamma h(\xi))$, then BSDE \eqref{eq4.2} becomes
	\begin{equation}\label{eq4.3}
		\y^{c, \xi}_{t}=g(\xi)-\int_{t}^{T}\gamma (\y^{c, \xi}_s)^{q} u(c_s)ds- \int_{t}^{T}\z^{ c, \xi}_{s}dB_s, ~~~t \in [0,T].
	\end{equation} 
	Since $g(\xi)\in(0,1)$ and the generator of BSDE \eqref{eq4.3} satisfies the monotonicity condition, 
	according to the Theorem 4.2 in Briand et al. \cite{briand2003lp}, there exists a unique solution in $\mathcal{S}^\beta \times \mathcal{M}^{\beta} $, $\beta>1$, for this BSDE. 
	
	As a result, finding the optimal strategy for \eqref{eq4.1} is equivalent to finding the optimal strategy of 
	\begin{equation}\label{eq4.4}
		\bar{V}(x)=\inf _{(c,\xi) \in \hat{\mathcal{A}}(x)} \bar{Y}_{0}^{c,\xi}, \quad x>0.
	\end{equation}
	Since the generator of \eqref{eq4.3} is decreasing and convex with respect to $c$, and the function $g(\cdot)$ is convex and decreasing in $\xi$, by the comparison theorem, i.e. Proposition 5 in Briand and Hu \cite{briand2006bsde}, we obtain that $(c,\xi) \rightarrow \bar{Y}_{0}^{c,\xi} $ is decreasing and convex.

	Furthermore, we need to study an optimization problem without constraints using theories of convex analysis.  For a fixed constant $v<0$, we now introduce the following optimization problem:
	\begin{equation}\label{eq4.5}
		\noindent \textbf{Auxiliary Problem:} ~~~~~~
		\inf _{(c,\xi) \in \mathcal{A}} \bar{J}(x, c, \xi, v),
	\end{equation}
	where the functional $\bar{J}$ is defined on $\mathcal{A}$ by
	\begin{align}\label{eq:barJ}
		\bar{J}(x, c, \xi, v)=\bar{Y}_{0}^{c, \xi}+v(x-X_0^{c,\xi}). 
	\end{align}
	
	\begin{proposition}\label{pro0}
		Suppose that assumptions \ref{a2}-\ref{a3-1} hold. For any given $x>0$,
		we have the following claims.
		\begin{itemize}
			\item[(i)]  There exists a constant $v^{*}=v^*(x)<0$, such that 
			\begin{equation}\label{eq4.6}
				\bar{V}(x)=\inf _{(c, \xi) \in \mathcal{A}} \bar{J}\left(x, c, \xi, v^{*}\right).
			\end{equation}
			\item[(ii)] If the infimum is attained in \eqref{eq4.4} by $(c^{*}, \xi^{*})$, then it achieves the infimum in \eqref{eq4.5} by $(c^{*}, \xi^{*})$  with $X_0^{c^*, \xi^*}=x$.
			\item[(iii)] Conversely, if there exists a constant  $v<0$ and it achieves the infimum in \eqref{eq4.5} by $(c^{*}, \xi^{*}) \in \mathcal{A}$  with $X_0^{c^*,\xi^*}=x$, then the infimum is also attained in \eqref{eq4.4} by $(c^{*}, \xi^{*})$.
		\end{itemize}
	\end{proposition}
\noindent{\textbf{Proof.}}
		We will prove the results by applying the classical convex result. Since $\mathcal{A}$ is convex, and Slater condition for the auxiliary problem optimization holds, for example, one can easily verify that $(c,\xi)=(\frac{x}{4T}, \frac{x}{4})\in \mathcal{A}$ and $X_0^{c,\xi}=\frac{x}{2}<x$.
		
		On the other hand,  for any constant $v<0$, one can easily find that
		\begin{align*}
			\bar{J}(x,c,\xi,v) &\geq \y^{c,\xi}_0+vx =\exp_q(-\gamma Y^{c,\xi}_0)+vx>vx>-\infty.
		\end{align*}
		Thus, equation \eqref{eq4.6} is finite. 
		Recalling Theorem 1 of Luenberger \cite[p. 217]{luenberger1997optimization}, we can prove (i) and (ii) directly. Besides, $v^*$ cannot be equal to zero since it reduces to a  optimization problem without constraints, and it yields to infinite optimal wealth or $\bar{V}(x)=-\infty$, which is impossible.
Finally, by Theorem 2 in \cite[p. 221]{luenberger1997optimization}, (iii) is obtained naturally. \hfill$\Box$
	
	By Proposition \ref{pro0}, similar to El Karoui, Peng, and Quenez \cite{el2001},  we only need to derive a necessary condition for the optimal solution of the optimization problem \eqref{eq4.5} by employing the fundamental properties of BSDEs in the subsequent analysis.
	
	\begin{theorem}\label{th2}(Maximum Principle)
		For $q>1$, $\gamma>0$ and $x>0$, suppose  assumptions \ref{a2}-\ref{a3-1} hold. Let $(c^{0}, \xi^{0}) \in \mathcal{A}$ be the optimal consumption and the optimal terminal wealth for problem \eqref{eq4.5} for a fixed deterministic constant $v<0$. Assume that $(Y^0,Z^0)=(Y^{c^0,\xi^0}, Z^{c^0,\xi^0})$ is the solution to BSDE \eqref{eq4.2}, or equivalently  $(\y^0, \z^0)=(\y^{c^0,\xi^0}, \z^{c^0,\xi^0})$ is the solution to BSDE \eqref{eq4.3}. Then  the following maximum principle holds
		\begin{align}
			- \gamma (\y^0_{0})^{q} (D^{0}_T)^q h'(\xi^0)&=v \tilde{D}_T ,~~ d\pp ~~ a.s.,\label{eq4.7}\\
			-\gamma (\y^0_0)^q (D^{0}_t)^q u'(c^0_t)&=v \tilde{D}_t, \quad t \in [0,T],~~ dt \times d\pp ~~ a.e.,\label{eq4.8}
		\end{align}
		where $$D^{0}=\mathcal{E}(\frac{\z^{0}}{\y^{0}}\cdot B)=\mathcal{E}\left(-\frac{\gamma  Z^0}{q \mu\left(Y^{0}\right)} \cdot B \right) \textrm{~~and~~}  \tilde{D}= \mathcal{E}(-\theta \cdot B)$$
	\end{theorem}
	
	\begin{remark}\label{max peng}
		The maximum principle (Theorem 4.2) in El Karoui, Peng, and Quenez \cite{el2001} requires that the utility functions satisfy some growth condition and the related BSDE's generator meets the uniformly Lipschitz condition in $(y,z)$. Compared with \cite{el2001}, our utility functions are more general and the structure of the BSDE is more complex. Our Theorem 4.1 can be also represented by the similar formulations as in \cite{el2001}, 
		\begin{align*}
			\Gamma_T^0 g'(\xi^0)&=v H^0_T ,~~ d\pp ~~ a.s.,\\
			\Gamma_t^0 f^0_c(t)&= vH_t^0,\quad t\in[0,T], ~~dt \times d\pp \quad a.e,
		\end{align*}
		where $\Gamma^0_t$ and $ H^0_t$, $t\in[0,T]$ are the adjoint processes given by \eqref{eq ga} and \eqref{eq H}, respectively, and $f^0_c(t)=-\gamma (\y_t^0)^q u'(c_t^0)$. It is also worth mentioning that our maximum principle is different from those obtained in Hu, Ji and Xu \cite{hu2022global} and Ji and Xu \cite{ji2024stochastic} for forward-backward stochastic control systems with quadratic generators.
	\end{remark}
	
	\begin{remark}
		When $q=1$, differently from the methodology of Faidi, Matoussi, and Mnif \cite{faidi2011maximization}, who considered the maximum principle of the 
		robust optimization via classical relative entropy, our method is still valid for the positive utility functions since we extend the stochastic flow approach in El Karoui, Peng, and Quenez \cite{el1997backward} beyond the Lipschitz condition.  More precisely, taking $\delta=0$, $\alpha=\bar{\alpha}=1$, $\gamma=\beta$, 
		$v^*=-\frac{v}{\gamma\bar{Y}_0^0}$, $q=1$, $\sigma=1$ and $M=B$, then our Theorem \ref{th2} is exactly Theorem 3.2 in \cite{faidi2011maximization}. 
	\end{remark}
	
	To prove the maximum principle (Theorem \ref{th2}), we adopt a different approach from that of Faidi, Matoussi, and Mnif \cite{faidi2011maximization}, whose methodology may fail to apply in our framework. Instead, we draw inspiration from the stochastic flow techniques developed by El Karoui, Peng, and Quenez \cite{el1997backward}. Specifically, we derive the key inequality \eqref{eq4.12} through analyzing the BSDE satisfied by the derivatives of processes $(\y^{\alpha}, \z^{\alpha})$, which are defined as follows.
	
	At first, let $(c,\xi) \in \mathcal{A}$ such that $\xi-\xi^0$, $c-c^0$ are uniformly bounded, i.e. $|\xi-\xi^0|+|c-c^0|\leq K$, $K$ is a constant. For any $\alpha \in [0,1]$,  let  $\xi^{\alpha}=\xi^0+\alpha(\xi-\xi^0)$ and $c^{\alpha}_t=c^0_t+\alpha(c_t-c^0_t)$, $t \in [0,T]$. Suppose the solution of related BSDE \eqref{eq4.3} associated with $(c^{\alpha}, \xi^{\alpha})$ is $(\y^{\alpha}, \z^{\alpha})$.  Motivated by Proposition 2.4 in El Karoui, Peng, and Quenez \cite{el1997backward}, one can  derive the auxiliary proposition below, whose proof is postponed to Appendix B.
	
	\begin{proposition}\label{pro1} Let the assumptions in Theorem $\ref{th2}$ hold, $p>q+1$,  and $(c^{\alpha}, \xi^{\alpha})$ and $(\y^{\alpha}, \z^{\alpha})$  are defined as above for all $\alpha\in[0,1]$. Then the function $\alpha \to (\y^{\alpha}, \z^{\alpha}) \in \mathcal{S}^p \times \mathcal{M}^p$ is differentiable, and its derivatives $(\partial_{\alpha} \y^{\alpha}, \partial_{\alpha} \z^{\alpha}) $ is  the unique solution in $\mathcal{S}^p \times \mathcal{M}^p$ of
		\begin{align*}
			\left\{\begin{array}{ll}
				-d \partial_{\alpha} \y^{\alpha}_{t} 
				&=\left[\partial_{y}f(\alpha, t,\y_{t}^{\alpha})\partial_{\alpha} \y^{\alpha}_{t}+\partial_{\alpha}f(\alpha, t,\y_{t}^{\alpha}) \right]dt-\partial_{\alpha} \z^{\alpha}_{t}dB_t, \\
				\partial_{\alpha} \y_{T}^{\alpha}&=\partial_{\alpha} g(\xi^{\alpha }),
			\end{array}\right.
		\end{align*}
		where  $f(\alpha, t, y)=-\gamma (y_+)^q u(c^{\alpha}_t)$ and $g(z)=\exp_q(-\gamma h(z)).$  
		
	\end{proposition}
	
	\begin{remark}
		It's worth noting that El Karoui, Peng, and Quenez \cite{el1997backward} utilized the stochastic flow method to obtain the BSDE satisfied by the derivatives $(\partial_{\alpha} Y^{\alpha}, \partial_{\alpha} Z^{\alpha}) $ of the function $\alpha \to (Y^{\alpha}, Z^{\alpha})$ under the uniformly Lipschitz condition. However, the generator of BSDE \eqref{eq4.2}
		is more intricate, thus we change our mindset to use the BSDE satisfied by $(\y,\z)$. Fortunately, the generator of BSDE \eqref{eq4.3} satisfies the monotonicity condition, although it doesn't fit into the setup of \cite{el1997backward}. By means of the a priori estimate of the solution, after overcoming some difficulties, we obtain the derivative of $\y^{\alpha}$ at $\alpha=0$ converges to $\partial_{\alpha} \y^{0}$ in $\mathcal{S}^p$ as $\alpha$ goes to zero. The details are fully elaborated in the proof.
	\end{remark}
	
	{\textbf{The proof of Theorem \ref{th2}}.}  
		From the above Proposition \ref{pro1}, $(\partial_{\alpha} \y^{0},\partial_{\alpha} \z^{0})$ is the solution to the BSDE as follows
		\begin{align*}
			\left\{\begin{array}{ll}
				-d \partial_{\alpha} \y^{0}_{t} 
				&=-\gamma q(\y^{0}_t)^{q-1} u(c^{0}_t)\partial_{\alpha} \y^{0}_{t}dt-\gamma(\y^{0}_t)^{q}u'(c^{0}_t)(c_t-c^0_t)dt-\partial_{\alpha} \z^{0}_{t}dB_t, \\
				\partial_{\alpha} \y_{T}^{0}&=g'(\xi^{0})(\xi-\xi^0).
			\end{array}\right.
		\end{align*}
		Also, we know $(\partial_{\alpha} X^{0},\partial_{\alpha} \pi^{0})$ is the solution of the following BSDE
		\begin{align*}
			\left\{\begin{array}{ll}
				-d \partial_{\alpha} X^{0}_{t} 
				&=((c_t-c^0_t)-\sigma_t \theta_t \partial_{\alpha} \pi^{0}_{t})dt-\partial_{\alpha} \pi^{0}_{t} dB_t, \\
				\partial_{\alpha} X_{T}^{0}&=\xi-\xi^0.
			\end{array}\right.
		\end{align*}
		
		Furthermore, in order to prove the maximum principle, we introduce the adjoint processes of $\partial_{\alpha}\y^{0}$ and $\partial_{\alpha}X^{0}$. Let $\Gamma^0$ represent the adjoint process associated with $\partial_{\alpha}\y^{0}$, defined as the solution to the SDE as follows
		\begin{align}\label{eq ga}
			\left\{\begin{array}{ll}
				d\Gamma^0_t&=-\Gamma_t^0\gamma q \bar{Y}_{t}^{q-1}u(c^0_t)dt, \quad t \in[0, T], \\
				\Gamma^0_0&=1.
			\end{array}\right.
		\end{align}
		Similarly, the adjoint process of $\partial_{\alpha}X^{0}$ is denoted by $H^0$, the solution of following SDE
		\begin{align}\label{eq H}
			\left\{\begin{array}{ll}
				dH_t^0&=-H_t^0 \theta_t dB_t, \quad t \in[0, T], \\
				H_0^0&=1.
			\end{array}\right.
		\end{align}
		
		Applying It\^{o}'s formula to $\Gamma^0_t \partial_{\alpha}\y^{0}_t$ and $H^0_t \partial_{\alpha}X^{0}_t$, respectively, it implies  that
		\begin{align}\label{eq4.13}
			\partial_{\alpha}\y^0_0-v\partial_{\alpha}X^0_0 
			=\E_{\pp} \left[(\Gamma^0_T g'(\xi ^0)-vH^0_T)(\xi-\xi^0)+\int_{0}^{T}(\Gamma^0_t f^0_c(t)-v H^0_t )(c_t-c^0_t)dt \right],
		\end{align}
        where $f^0_c(t)=-\gamma (\y_t^0)^q u'(c_t^0)$.

		Since $(c^0,\xi^0)$ is the optimal strategy of problem \eqref{eq4.5}, then for any $\alpha \in [0,1]$,
		$$\y^{\alpha}_0-vX^{\alpha}_0 \geq \y^0_0-vX^0_0 .$$
		Dividing $\alpha$ by the above inequality and letting $\alpha$ tend to 0, we get
		\begin{align}\label{eq4.12}
			\partial_{\alpha}\y^0_0-v\partial_{\alpha}X^0_0 \geq 0,
		\end{align}
		where $\partial_{\alpha}\y^0_0$ and $\partial_{\alpha}X^0_0$ denote the (right) derivatives of $\y^{\alpha}_0$ and $X^{\alpha}_0$ at $\alpha=0$ respectively.
		
		Combining inequality \eqref{eq4.12} , recalling the arbitrariness of  $(c,\xi) \in \mathcal{A}$ such that $\xi-\xi^0$ and $c-c^0$ are bounded, then we have 
		\begin{align}\label{4.14}
			\E_{\pp}\left[(\Gamma^0_T g'(\xi ^0)-vH^0_T)(\xi-\xi^0) \right] \geq 0
		\end{align}and  
		$$\E_{\pp}\left[\int_{0}^{T}(\Gamma^0_t f^0_c(t)-v H^0_t )(c_t-c^0_t)dt \right] \geq 0.$$
		
		On one hand, let $A=\left\{\Gamma^0_T g'(\xi ^0)-vH^0_T <0 \right\}$,  $\xi:=\xi^0+\textbf{1}_A$, then one can easily verify that $\E_{\pp}[(h'(\xi))^{2p}]\leq \gamma^p\E_{\pp}[(h'(\xi^0)^{2p}]<+\infty$, which implies $(c^0, \xi)\in\mathcal{A}$.
		Taking account of $\eqref{4.14}$,  one has  $\pp(A)=0$, which means that
		\begin{align}\label{geq}
			\Gamma^0_T g'(\xi ^0)-vH^0_T \geq 0, ~~d\pp~~ a.s.
		\end{align}
		As Inada condition $h'(0)=+\infty$ holds, that is, $g'(0)=- \gamma \exp^q_{q}(-\gamma h(0))h'(0)=-\infty$. Hence,  it ensures that $\xi^0>0 ~a.s.$ by $\eqref{geq}$.
		
		In the following, we claim  that  
		\begin{align}\label{eq4.15}
			\Gamma^0_T g'(\xi ^0)-vH^0_T = 0, ~~d\pp~~ a.s.
		\end{align}
		holds on $\{\xi^0>0\}$. Indeed, for any $ \epsilon >0$, considering the set  $$B_{\epsilon}=\left\{\Gamma^0_T g'(\xi ^0)-vH^0_T >0, ~~2\epsilon < \xi^0 < \frac{1}{\epsilon} \right\},$$  and taking $\xi:=\xi^0-\epsilon \textbf{1}_{B_{\epsilon}}$, i.e. $\xi=\xi^0 \textbf{1}_{B^c_{\epsilon}}+(\xi^0-\epsilon)\textbf{1}_{B_{\epsilon}}$, where $B^c_{\epsilon}$ denotes the complement of $B_{\epsilon}$, we have that
		$\E_{\pp}[|h'(\xi)|^{2p}]=\E_{\pp}[|h'(\xi^0)|^{2p}\textbf{1}_{B^c_{\epsilon}}+|h'(\xi^0-\epsilon)|^{2p}\textbf{1}_{B_{\epsilon}}]<+\infty.$ Combined with $\eqref{4.14}$, we can obtain that $\pp(B_{\epsilon})=0$. Sending $\epsilon \to 0$, one finally gets $\pp(\Gamma^0_T g'(\xi ^0)-vH^0_T >0)=0$. Therefore, \eqref{eq4.15}
		holds. 
		
		On the other hand, according to the SDEs for the adjoint processes, we have
		$$\Gamma^0_T=\exp\left(-\int_{0}^{T}\gamma q (\bar{Y}^0_{s})^{q-1}u(c^0_s)ds\right) .$$
		Besides, we know
		\begin{align*}
			g'(\xi^0) &=- \gamma \exp^{q}_q(-\gamma h(\xi^0))h'(\xi^0) =- \gamma (\bar{Y}^0_{T})^{q}h'(\xi^0),
		\end{align*}
		and
		$$\bar{Y}^0_{T}=\bar{Y}^0_{0}\exp\left(\int_{0}^{T}\gamma (\bar{Y}^0_{s})^{q-1} u\left(c^0_{s}\right)ds\right)\mathcal{E}\left(\frac{\bar{Z}^0}{\bar{Y}^0}\cdot B\right)_T .$$
		The same is for $H^0_T$, so we can finally obtain the necessary condition for the optimal terminal wealth $\xi^0$ as follows
		$$- \gamma (\y^0_{0})^{q} (D^{0}_T)^q h^{'}(\xi^0)=v \tilde{D}_T ,~~ d\pp ~~ a.s.$$
		Therefore, the equality $\eqref{eq4.7}$ is proved. Similarly, the necessary condition to be satisfied for optimal consumption $c^0$ can also be derived. \hfill$\Box$

	\section{Existence of the optimal strategy and the coupled forward-backward system}\label{sec:5}

	This section investigates the existence of optimal solution to the optimization problem and establishes the associated forward-backward system based on the maximum principle to characterize the optimal solution.
    
\subsection{Existence of an optimal strategy}\label{subsec:5.1}
This subsection is primarily devoted to establishing the existence of an optimal strategy for the optimization problem \eqref{eq4.1}. 
We first introduce the following compactness assumption on a regular minimizing sequence. 


\begin{assumption}\label{a4}
For every initial capital $x>0$, there exists a minimizing
sequence $(c^n,\xi^n)_{n\geq 1}\subset\hat{\mathcal{A}}(x)$ for \eqref{eq4.4} such that
$$\bar Y_0^{c^n,\xi^n}\longrightarrow\bar V(x) \text{~~and~~}
\sup_{(c,\xi)\in\operatorname{co}\{(c^n,\xi^n)_{n\ge1}\}}\mathfrak M(c,\xi)<+\infty,
$$where $\mathfrak M(c,\xi)=\mathbb E_{\pp}[\int_0^T\bigl(c_t^\beta+u'(c_t)^{2p}\bigr)dt+
\xi^\beta+h'(\xi)^{2p}]$.
Here, $\operatorname{co}$ denotes the set of finite convex combinations with
deterministic nonnegative coefficients summing to one. 
\end{assumption}

Assumption \ref{a4} is a compactness and uniformly integrability condition rather than a very restrictive assumption. It ensures that a minimizing sequence for \eqref{eq4.4} can be chosen so that the relevant powers of consumption, terminal wealth, and marginal utilities remain uniformly integrable; this is exactly what is needed for the lower semicontinuity argument and the use of Mazur's lemma in the proof of Theorem \ref{th exist}. The assumption is satisfied in many standard models. For instance, let $u(c)=h(c)=c^{a}/a$ with $(0<a<1)$, and suppose that the market price of risk is bounded. Then the density $\tilde D=\mathcal{E}(-\theta\cdot B)$ has moments of all positive and negative orders. The associated optimal pair is given by
$c_t^*=(y\tilde D_t)^{-1/(1-a)}$ and 
$\xi^*=(y\tilde D_T)^{-1/(1-a)}$,
where $y>0$ is chosen so that the budget constraint
$X_0^{c^*,\xi^*}=x$
holds. Since $\tilde D$ has all positive and negative moments, one has $\mathfrak{M}(c^*,\xi^*)<+\infty$ under the standing parameter restrictions. Taking the constant minimizing sequence, we obtain
Assumption \ref{a4} holds automatically in this power utility example, showing that it is a mild technical condition.

\begin{theorem}\label{th exist}
Suppose the assumptions A1-A4 hold. For every initial
capital $x>0$ for which A4 holds, there is a unique $(c^*,\xi^*)\in \hat{\mathcal{A}}(x)$ that attains the optimal value of problem \eqref{eq4.1}, equivalently in \eqref{eq4.4}. Moreover, $X_0^{c^*,\xi^*}=x$.  
\end{theorem}
	
By transforming the constrained maximization problem \eqref{eq4.1} into an equivalent minimization problem \eqref{eq4.4}, and subsequently converting it into an unconstrained formulation, we can firstly get the existence of a solution to problem \eqref{eq4.4} by the unconstrained problem \eqref{eq4.5} with fixed $v$. The proof combines convex combinations of a minimizing sequence with lower
semicontinuity and is given at the end of Appendix C. The following lemmas
establish the required convergence properties, and their proofs are also given
in Appendix C.

\begin{lemma}\label{lem2}
Let $(c^n,\xi^n)\in \mathcal{A}$ converge to $(c,\xi)$ in measure, under $d\pp\times dt$ for
consumption and under $\mathbb P$ for terminal wealth. If$$\sup_n\mathfrak M(c^n,\xi^n)<+\infty,$$
then $(c,\xi)\in \mathcal{A}$ and $X_0^{c^n,\xi^n}\to X_0^{c,\xi}$. In particular,
if $(c^n,\xi^n)\in \hat{\mathcal{A}}(x)$ for every $n$, then
$(c,\xi)\in\hat{\mathcal{A}}(x)$. The families
$$
\{u(c^n)^p\}_{n\geq1},\quad\{u'(c^n)^p\}_{n\geq1},\quad
\{h(\xi^n)^p\}_{n\geq1},\quad\{h'(\xi^n)^p\}_{n\geq1}$$
are uniformly integrable on their respective measure spaces. 
\end{lemma}
\begin{lemma}\label{lem3}
For each $x>0$ and $v<0$, the functional $\bar J(x,\cdot,\cdot,v)$ is
strictly convex on $\mathcal{A}$. For every sequence satisfying the hypotheses of
Lemma \ref{lem2}, then
$$
\bar Y_0^{c,\xi}\le\liminf_n\bar Y_0^{c^n,\xi^n} \text{~~and~~}
\bar J(x,c,\xi,v)\le\liminf_n\bar J(x,c^n,\xi^n,v).
$$  
\end{lemma}

Building upon the aforementioned properties, we establish the existence of a solution to problem \eqref{eq4.5} via the following theorem.
\begin{theorem}\label{th noconstraint}
Under the assumptions of Theorem \ref{th exist}, let $v^*=v^*(x)<0$ be a multiplier
associated with the optimizer at initial capital $x$, as in Proposition \ref{pro0}.
Then,  the auxiliary problem \eqref{eq4.5},  with this fixed $v^*$, has a unique
minimizer $(c^*,\xi^*)$ in $\mathcal{A}$.  
\end{theorem}
{\textbf{Proof.}}
Theorem \ref{th exist} establishes attainment of the budget optimum. Proposition \ref{pro0}
therefore implies that $(c^*,\xi^*)$ minimizes \eqref{eq4.5} with the associated
multiplier. Strict convexity in Lemma \ref{lem3} yields uniqueness.
\hfill$\Box$

\begin{remark}
Employing functional-analytic approaches, El Karoui, Peng and Quenez \cite{el2001} and Faidi, Matoussi, and Mnif \cite{faidi2011maximization} respectively establish the existence of solutions to the optimization problem in different environments. Reference \cite{el2001}  penalizes the objective functional to induce strong convexity, thereby obtaining the existence of an optimal strategy for the penalized problem, and subsequently proving that the sequence of optimal strategy is uniformly bounded in the weak topology, which ultimately yields the existence of an optimal strategy for the original problem.  \cite{faidi2011maximization} principally employs convex analysis techniques, where they assume exponential boundedness ensuring power-integrability of both consumption and terminal wealth, consequently guaranteeing the sequence's convergence and uniform integrability in the specified space. 
Under an explicit compactness assumption on a regular minimizing sequence, we prove existence and
uniqueness of the optimal strategy. 
\end{remark}
	
We end this subsection by a dual identity.  
We introduce the  functional $\tilde{V}$ by 
$$\tilde{V}(v)= \inf _{(c,\xi) \in \mathcal{A}}\{\y_0^{c,\xi}-v X_0^{c,\xi} \},~~~~v<0.$$

For every $(c,\xi)\in\hat{\mathcal{A}}(x)$,
$$\tilde V(v)+vx\le\bar Y_0^{c,\xi}+v(x-X_0^{c,\xi})
\le\bar Y_0^{c,\xi}.$$
Taking the infimum over admissible pairs establishes weak duality.
At the multiplier $v^*$ of Theorem \ref{th noconstraint}, Proposition \ref{pro0} and
$X_0^{c^*,\xi^*}=x$ yield equality. Consequently,
\begin{equation}\label{eq5.5}
\bar V(x)=\sup_{v<0}\{\tilde V(v)+vx\}
=\tilde V(v^*)+v^*x.
\end{equation}

		\subsection{Forward-backward system}

		In this subsection, we establish a characterization of the optimal consumption investment strategy $(c^*,\xi^*)$ via a forward-backward system, as a direct consequence of the maximum principle.
		
		Let us denote $(\Gamma^0, H^0)$ and $(\y^0,\z^0)$ by $(\Gamma^*, H^*)$ and $(\y^*,\z^*)$ respectively. In fact, from Theorem \ref{th2} and Theorem $\ref{th exist}$, it follows that the optimal consumption $c^*_t$, $t \in [0,T]$ and terminal wealth $\xi^*$ are given by
		\begin{align}
			c^*_t&=I_1 \left(-\frac{v}{\gamma}   (\y^*_0 D^{*}_t)^{-q} \tilde{D}_t \right)=I_1 \left( -\frac{v}{\gamma} H^*_t (\y^*_t)^{-q} (\Gamma^*_t)^{-1}  \right),~t \in [0,T],~ dt \times d\pp~ a.e., \label{optimal}\\
			\xi^*&=I_2 \left(-\frac{v}{\gamma} (\y^*_{0} D^{*}_T)^{-q} \tilde{D}_T \right)=I_3 \left(v  H^*_T (\Gamma^*_T)^{-1} \right), ~d\pp ~a.s.\label{eq optimal}
		\end{align}
		where $I_1$ (resp., $I_2$, $I_3$) is equal to $(u')^{-1}(\cdot)$ (resp., $(h')^{-1}(\cdot)$, $(g')^{-1}(\cdot)$), i.e., the inverse of derivative function of $u$ (resp., $h$, $g$). The following result is easily from Theorem \ref{th2} and Theorem \ref{th exist}.
		
\begin{theorem}
Under the assumptions of Theorem \ref{th exist}, the optimal strategy at capital $x$ is characterized by the following system. The multiplier is an unknown deterministic constant $v<0$.  Set $\bar Y=\exp_q(-\gamma Y)$, $\bar Z=-\gamma\bar Y^qZ$, and define the strategy by the feedback formulas
$$
c_t=I_1\left(\frac{-vH_t^*}{\gamma\Gamma_t^*(\bar Y_t)^q}\right) \text{~~and~~}
\xi=I_3\left(\frac{vH_T^*}{\Gamma_T^*}\right).
$$
The coupled equations are
\begin{align*}
\begin{cases}
dX_t=(\pi_t\sigma_t\theta_t-c_t)dt+\pi_t\sigma_t\,dB_t,
&X_0=x,\quad X_T=\xi,\\
dY_t=\left[\dfrac\gamma2\dfrac{|Z_t|^2}{\mu(Y_t)}-u(c_t)\right]dt+Z_t\cdot dB_t,
&Y_T=h(\xi),\\
d\Gamma_t^*=-\gamma q\Gamma_t^*(\bar Y_t)^{q-1}u(c_t)dt,
&\Gamma_0^*=1,\\
dH_t^*=-H_t^*\theta_t\cdot dB_t,
&H_0^*=1.
\end{cases}
\end{align*}
The admissible solution class includes that $Y$ is continuous and nonnegative,  $\bar{Y}$ is of class $D$, and $\int_0^TZ_s^2ds<+\infty$, 
$(X,\pi\sigma)\in \mathcal{S}^\beta\times \mathcal{M}^\beta$ and $(\bar Y,\bar Z)\in \mathcal{S}^p\times \mathcal{M}^p $
with $(c,\xi)\in \mathcal{A}$. The density of its minimizing model is
$$
D_t^*=\mathcal E\left(\frac{\bar Z}{\bar Y}\cdot B\right)_t
=\frac{\bar Y_t}{\bar Y_0}\exp\left(-\gamma\int_0^t(\bar Y_s)^{q-1}u(c_s)ds\right).
$$The system has a unique solution in this class, including its multiplier. 
\end{theorem}
{\textbf{Proof.}} Theorems \ref{th exist} and \ref{th noconstraint}, together with the maximum principle, yield a solution
consisting of the optimal strategy, their replicating wealth, and the
associated utility and adjoint processes. Conversely, let a solution in the
stated class be given. From the first equation, one has that $X_0^{c,\xi}=x$, while Theorem \ref{th1}
identifies its utility and its minimizing model $\Q^*$, whose density $D^*$
is bounded and strictly positive. The feedback formulas imply
$$(D_t^*)^qu'(c_t)=y\tilde D_t,\qquad
(D_T^*)^qh'(\xi)=y\tilde D_T,\qquad
y=\frac{-v}{\gamma(\bar Y_0)^q}>0,
$$where, $\tilde{D}=H^*$.

For any $\Q\in\mathcal{Q}_f$ and $(\tilde c,\tilde \xi)\in\hat{\mathcal{A}}(x)$, define
$$F(\Q;\tilde c,\tilde \xi):=\mathbb E_{\pp}\left[(D_T^\Q)^qh(\tilde\xi)
+\int_0^T(D_t^\Q)^qu(\tilde c_t)dt\right]
+\frac1\gamma H_q(\Q\mid\mathbb \pp).$$
By Theorem \ref{th1}, one has that
$Y_0^{\tilde c,\tilde \xi}=\inf_{\Q\in \mathcal{Q}_f}F(\Q;\tilde c,\tilde \xi).$
Therefore, for any $(\tilde c,\tilde \xi)\in\hat{\mathcal{A}}(x)$, concavity of the
utilities and the budget representation imply that 
\begin{align*}
Y_0^{\tilde c,\tilde \xi}
&\le F(\Q^*;\tilde c,\tilde\xi)\\
&\le F(\Q^*;c,\xi)
+y\bigl(X_0^{\tilde c,\tilde\xi}-X_0^{c,\xi}\bigr)
\le F(\Q^*;c,\xi)=Y_0^{c,\xi}.
\end{align*}
Thus, the candidate strategy $(c,\xi)$ is optimal, and it is unique by the strict concavity (resp.  convexity) of $F(\Q;\cdot,\cdot)$ (resp. $\bar J(x,\cdot,\cdot,v)$) on $\mathcal{A}$. Then, the wealth equation, BSDEs and the adjoint equations determines all the processes. Finally,
one gets that the multiplier is uniquely determined by 
\begin{align*}
 v=-y\gamma (\bar{Y}_0)^q=-\frac{\gamma (\bar{Y}_0)^q(D_T^*)^qh'(\xi)}{\tilde{D}_T}= \Gamma_T^*g'(\xi)/\tilde D_T,
\end{align*}where one uses the facts that $g'(\xi)=-\gamma (\bar{Y}_T)^q h'(\xi)$ and $(D_T^*)^q=(\bar{Y}_T)^q\Gamma_T^*/(\bar{Y}_0)^q$. \hfill$\Box$

		\begin{remark}
			Different from El Karoui, Peng, and Quenez \cite{el2001} where the Lipschitz condition is critical, we investigate a quadratic BSDE under monotonicity conditions, with all processes being explicitly constructed.	Furthermore, compared with Faidi, Matoussi, and Mnif \cite{faidi2011maximization}, we obtain a more specialized BSDE with quadratic growth in $Z$, while additionally accounting for the influence of adjoint processes.
		\end{remark}
		
		\begin{example}
			If there is no consumption item, 
			the recursive relation is
			$$Y_0^{\xi}=-\frac{1}{\gamma}\ln_{q}\E_{\pp}[\exp_{q}(-\gamma h(\xi))].$$
			Then the optimal problem \eqref{eq4.1} is related to the following problem:
			\begin{align*}
				V^{Ex}(x):=\sup_{\xi \in \mathcal{B}(x)}\E_{\pp}[-\exp_{q}(-\gamma h(\xi))],
			\end{align*}
			where $\mathcal{B}(x)=\{\xi \in L_{+}^{\beta}(\F_T)|~\E_{\pp}[h'(\xi)^{2p}]< +\infty ~and~X_0 ^{\xi} \leq x \}$.
			The utility function $U^{Ex}(z)=-\exp_{q}(-\gamma h(z))$ is concave and increasing, which satisfies the Inada condition. Suppose assumption A4 hold for terminal wealth.   
            By the Theorem 2.0 (Complete case) in Kramkov and Schachermayer \cite{kramkov1999asymptotic}, the optimal terminal wealth is given by
			\begin{align}\label{eq KS}
				\xi^*=I^{Ex}(y \tilde{D}_T),
			\end{align}
			where $I^{Ex}(z)=((U^{Ex})')^{-1}(z)$, $y=(V^{Ex})'(x)$. Taking $h(x)=x^a/a$.  One can verify that
$(U^{Ex})'(z)$ behaves like $z^{a-1}$ for small $z>0$.
For large $\tilde D_T$ (the only dangerous region where $\xi^*\to0$ and $h'(\xi^*)$ may blow up), we obtain a tail bound $h'(\xi^*) = O(\tilde D_T)$. For small $\tilde D_T$, $\xi^*$ stays away from zero so $h'(\xi^*)$ is deterministically bounded. Splitting the expectation over these two regimes and using that $\tilde D_T$ admits all finite moments due to bounded $\theta$ yields the finiteness of $\mathbb E_{\pp}[h'(\xi^*)^{2p}]$. Besides, one can similar prove that $\mathbb E_{\pp}[(\xi^*)^{k}]<+\infty$ for any $k>0$.

			From \eqref{eq optimal} and \eqref{eq KS}, we can get
			\begin{align}\label{eq inverse}
				I_2 \left(-\frac{v}{\gamma} (\y^*_{0} D^{*}_T)^{-q} \tilde{D}_T \right)=I_3 \left(v H^*_T (\Gamma^*_T)^{-1} \right)=I^{Ex}(y \tilde{D}_T).
			\end{align}
			Since there is no consumption, we have $v H^*_T (\Gamma^*_T)^{-1}=v\tilde{D}_T$. Combined with $g(z)=-U^{Ex}(z)$, we know that $I^{Ex}(z)=((-g)')^{-1}(z)$ and equation \eqref{eq inverse} can be rewritten as
			$$(g')^{-1} \left(v \tilde{D}_T \right)=((-g)')^{-1}(y \tilde{D}_T).$$
			So we derive that $v=-y$
			and
			$$D^*_T=\left[\frac{- \gamma (\y^*_{0})^q h' \left(I^{Ex}(y \tilde{D}_T) \right)}{v \tilde{D}_T} \right]^q.$$
			As a result, we obtain that
			$$V(x)=Y_0^{\xi^*}=-\frac{1}{\gamma}\ln_{q}\E_{\pp}[\exp_{q}(-\gamma h(\xi^*))].$$
		\end{example}

 \section*{Acknowledgments}

The authors thank the grants supported by the National Natural Science Foundation of China (No.12671583, No. 12671580).


\appendix

\section*{Appendix: The Proofs}

\section*{A. The equivalent robust optimization problem}
\addcontentsline{toc}{section}{A. The equivalent robust optimization problem}
\renewcommand{\theequation}{A.\arabic{equation}}
\setcounter{equation}{0}
\renewcommand{\thelemma}{A.\arabic{lemma}}
\setcounter{lemma}{0} 

We explain here that the corresponding robust portfolio choice problem. For initial wealth $y\ge0$, define
$$
\bar{\mathcal{B}}(y)=\left\{(c,\pi):
c\in \mathcal{C}^\beta, \pi\sigma\in \mathcal{M}^\beta,\
X^{y,c,\pi}\ge0, (c,X_T^{y,c,\pi}) \in \mathcal{A}\right\},
$$
where $dX_t^{y,c,\pi}=(\pi_t\sigma_t\theta_t-c_t)dt+
\pi_t\sigma_tdB_t$, $X_0^{y,c,\pi}=y.$ 
Since the wealth equation is affine in
$(c,\pi)$, one obtains the set $\bar{\mathcal{B}}(y)$ is convex.  For a forward strategy, set $Y_0^{y,c,\pi}=Y_0^{c,X_T^{y,c,\pi}}$.

More precisely, for $x>0$, we show that 
			\begin{align}\label{forward}
				\sup_{(c,\pi) \in \bar{\mathcal{B}}(x)}Y_0^{x,c,\pi}=\sup_{0\leq y \leq x} \sup_{(c,\pi) \in \bar{\mathcal{B}}(y)} Y_0^{y,c,\pi}
			\end{align}and identify this common value with
			\begin{align}\label{backward}
				\sup_{(c,\xi) \in \hat{\mathcal{A}}(x)}Y_0^{c,\xi},
			\end{align}
            where $Y_0^{c,\xi}$  is given by BSDE \eqref{eq3.1} with $\zeta=h(\xi)$ and $U_\cdot=u(c_\cdot)$.  
            
First, one shows the identity of \eqref{forward}. In fact, it can be easily obtained by the comparison theorem for $\check{Y}_0^{c,X_T^{y,c,\pi}}=\exp_q(-rY_0^{c,X_T^{y,c,\pi}})$ with respect to $y$.			
			
Next, we verify the relationship between \eqref{forward} and  \eqref{backward}. Indeed, for any $0\leq y\leq x$, suppose that $(c,\pi) \in \bar{\mathcal{B}}(y)$ and $X_T^{y,c,\pi}=\xi$, then along this path evaluating the wealth process at the initial time yields $X_0^{c,\xi}=y \leq x$. Thus, we have $(c,\xi) \in \hat{\mathcal{A}}(x)$, and 
			\begin{align}\label{}
				\sup_{0\leq y \leq x} \sup_{(c,\pi) \in \bar{\mathcal{B}}(y)} Y_0^{y,c,\pi} \leq \sup_{(c,\xi) \in \hat{\mathcal{A}}(x)}Y_0^{c,\xi}.
			\end{align}
			
			Conversely, if $(c,\xi) \in \hat{\mathcal{A}}(x)$, then there exists some $\pi$, s.t. $X_T^{y,c,\pi}=\xi$, then $(c,\pi) \in \bar{\mathcal{B}}(y) $, and we have
			\begin{align}\label{}
				\sup_{(c,\xi) \in \hat{\mathcal{A}}(x)}Y_0^{c,\xi}  \leq \sup_{0\leq y \leq x} \sup_{(c,\pi) \in \bar{\mathcal{B}}(y)} Y_0^{y,c,\pi}. 
			\end{align}

Therefore, we can get the problem \eqref{forward} is equivalent to \eqref{backward}. We will focus on the BSDE \eqref{eq wealth} and restrict our analysis to the optimization problem related to $(c,\xi)$, i.e.,
			$V(x)=\sup_{(c,\xi) \in \hat{\mathcal{A}}(x)}Y_0^{c,\xi}$.
			
\section*{B.  Proof of Proposition \ref{pro1}}
\addcontentsline{toc}{section}{B. Proof of Proposition \ref{pro1}}
\renewcommand{\theequation}{B.\arabic{equation}}
\setcounter{equation}{0}
\renewcommand{\thelemma}{B.\arabic{lemma}}
\setcounter{lemma}{0}

We will prove Proposition \ref{pro1} in the following two steps.
			
				\textbf{Step 1:} Firstly, we need to prove that for some $p>q+1$, 
				$$\lim_{\alpha \to 0}	\E_{\pp} \left[\underset{0\leq t \leq T}{\sup}|\y^{\alpha}_t-\y^{0}_t|^p \right]=0.$$
				Since the terminal value of BSDE \eqref{eq4.3} is bounded and $u(\cdot)$ is integrable, then according to the Proposition 3.2 of Briand et al. \cite{briand2003lp}, one can obtain that for any $k>1$,
				\begin{align*}
					&\E_{\pp} \left[\underset{0\leq t \leq T}{\sup}|\y^{\alpha}_t-\y^{0}_t|^k \right]+\E_{\pp} \left[ \left(\int_{0}^{T}|\z^{\alpha}_s-\z^{0}_s|^{2} ds\right)^{\frac{k}{2}} \right] \\
					\leq &C_k \left(\E_{\pp} \left[|g(\xi^{\alpha})-g(\xi^{0})|^k \right] + \E_{\pp} \left[\int_{0}^{T}|\y^{\alpha}_s-\y^{0}_s|^{k-1}|\y^{0}_s|^q |u(c^{\alpha}_s)-u(c^{0}_s)|ds \right] \right),
				\end{align*} 
				where $C_k$ is a constant depending only on $k$.
				Furthermore, by Young's inequality, for every $\epsilon>0$, 
				\begin{align*}
					&\E_{\pp} \left[\int_{0}^{T}|\y^{\alpha}_s-\y^{0}_s|^{k-1}|\y^{0}_s|^q |u(c^{\alpha}_s)-u(c^{0}_s)|ds \right] \\
					\leq & \epsilon \E_{\pp} \left[\underset{0\leq t \leq T}{\sup}|\y^{\alpha}_t-\y^{0}_t|^k \right] 
					+C_{k,\epsilon} \E_{\pp} \left[ \left(\int_{0}^{T}|\y^{0}_s|^q |u(c^{\alpha}_s)-u(c^{0}_s)|ds \right)^k \right].
				\end{align*}
Since $0<\bar Y^0\le1$, the last expectation is bounded by $T^{k-1}\mathbb E\int_0^T
|u(c^\alpha_s)-u(c^0_s)|^kds.$  Take $k=p$ and choose $\varepsilon$ so that $C_p\varepsilon<1$,
where $C_p$ is the constant in the preceding stability estimate.

We already know that $\xi-\xi^0$, $c-c^0$ are uniformly bounded, and  $u(\cdot)$ is an increasing concave function, $c_{.} \geq 0$. Thus
				\begin{align*}
					u(c^{\alpha})-u(c^{0}) \geq -u(c^0)
				\end{align*}
				and
				\begin{align*}
					u(c^{\alpha})-u(c^{0})=u(c^0+\alpha(c-c^0))-u(c^0)\leq u(c^0+K)-u(c^0)\leq K u'(c^0).
				\end{align*}
				Similarly, $g(\xi) \geq 0$ is convex and decreasing, so we have 
				\begin{align*}
					g(\xi^0+K)-g(\xi^0) \geq Kg'(\xi^0)
				\end{align*}
				and
				\begin{align*}
					g(\xi^0+K)-g(\xi^0) 
					&\leq g(\xi^{\alpha})-g(\xi^{0}) \leq g(\xi^0)+g(\xi).
				\end{align*}
				Consequently, $|u(c^{\alpha})-u(c^{0})| \leq  K u'(c^0)+ u(c^0)$, $|g(\xi^{\alpha})-g(\xi^{0})| \leq g(\xi^0)+g(\xi)+K|g'(\xi^0)|$.
				Hence,  $u(c^0), u'(c^0), g(\xi), g(\xi^0)$ and $g'(\xi)$ are both $p$-order integrable by assumption A3 and definition $\mathcal{A}$.
				
				Noting that when $\alpha \to 0$, $u(c^{\alpha}) \to u(c^0)$, and $g(\xi^{\alpha}) \to g(\xi^{0})$ in measure space.  Then by the dominated convergence theorem, for the $p>q+1$, we can obtain that
				$$\lim_{\alpha \to 0}\E_{\pp}[\sup_{0\leq t \leq T}|\y^{\alpha}_t-\y^{0}_t|^p ]+\E_{\pp}\Big[\big(\int_{0}^{T}|\z^{\alpha}_s-\z^{0}_s|^{2} ds\big)^{\frac{p}{2}} \Big]=0.$$

				\textbf{Step 2:}
				Similar to Proposition 2.4 in El Karoui, Peng, and Quenez \cite{el1997backward}, when $\alpha \to 0$, we show that $(\Delta_{\alpha} \y, \Delta_{\alpha} \z)$ converges to $(\partial_{\alpha}\y^0, \partial_{\alpha}\z^0)$ in $\mathcal{S}^p \times \mathcal{M}^p$, as $\alpha$ tends to $0$.

				We know that  $(\partial_{\alpha} \y^{0}, \partial_{\alpha} \z^{0})$ is the solution of the following BSDE:
				\begin{align*}
					\left\{\begin{array}{ll}
						-d \partial_{\alpha} \y^{0}_{t} 
						&=-\gamma q(\y^{0}_t)^{q-1} u(c^{0}_t)\partial_{\alpha} \y^{0}_{t}dt-\gamma(\y^{0}_t)^{q}u'(c^{0}_t)(c_t-c^0_t)dt-\partial_{\alpha} \z^{0}_{t}dB_t, \\
						\partial_{\alpha} \y_{T}^{0}&=g'(\xi^{0})(\xi-\xi^0),
					\end{array}\right.
				\end{align*}
				where $\partial_{y}f(0, t,\y_{t}^{0})=-\gamma q(\y^{0}_t)^{q-1} u(c^{0}_t)$, $\partial_{\alpha}f(0, t,\y_{t}^{0})=-\gamma(\y^{0}_t)^{q}u'(c^{0}_t)(c_t-c^0_t)$.
				
				Let $\Delta_{\alpha} \bar{Y}_{t}=\frac{\y_t^{\alpha}-\y_t^{0}}{\alpha}$, $\Delta_{\alpha} \z_{t}=\frac{\z_t^{\alpha}-\z_t^{0}}{\alpha}$, we have BSDE as follows
				\begin{align*}
					\left\{\begin{array}{ll}
						-d \Delta_{\alpha} \y_{t} &=\alpha^{-1} \left[f(\alpha, t,\y_{t}^{\alpha})-f(0, t,\y_{t}^{0}) \right]dt-\Delta_{\alpha} \z_{t}dB_t \\
						&=-\alpha^{-1} \left[\gamma(\y^{\alpha}_t)^q u(c^{\alpha}_t)-\gamma(\y^{0}_t)^q u(c^{0}_t) \right]dt-\Delta_{\alpha} \z_{t}dB_t,\\
						\Delta_{\alpha} \y_{T} & =\alpha^{-1} \left(g(\xi^{\alpha})-g(\xi^{0}) \right).
					\end{array}\right.
				\end{align*}
				Rewriting the first equation, one derives that
				$$-d \Delta_{\alpha} \y_{t}=[A_{\alpha}(t)\Delta_{\alpha} \y_{t}+\varphi_{\alpha}(t)]dt-\Delta_{\alpha} \z_{t}dB_t,$$
				where
				$$		A_{\alpha}(t)=
				\left\{\begin{array}{ll}
					\frac{f(\alpha, t, \y_t^{\alpha})-f(\alpha, t, \y_t^{0} )}{\y_t^{\alpha}-\y_t^{0}}=\frac{-\gamma u(c^{\alpha}_t)[(\y^{\alpha}_t)^q-(\y^{0}_t)^q]}{\y_t^{\alpha}-\y_t^{0}}, \quad \y_t^{\alpha}\neq \y_t^{0},\\
					\partial_{y}f(\alpha, t, \y_t^{0}), \quad \text{otherwise},
				\end{array}\right.$$
				$$ \varphi_{\alpha}(t)=\frac{1}{\alpha}(f(\alpha, t, \y^0_t)-f(0,t,\y^0_t))=\frac{-\gamma (\y^{0}_t)^q[u(c^{\alpha}_t)- u(c^{0}_t)]}{\alpha}.$$
				From step 1 we have proved $(\y^{\alpha}, \z^{\alpha})$ converges to $(\y^0, \z^0)$ in $\mathcal{S}^p \times \mathcal{M}^p $, and now we need to show that, as $\alpha$ goes to $0$,  $(\Delta_{\alpha} \y, \Delta_{\alpha} \z)$ converges to  $(\partial_{\alpha}\y^0, \partial_{\alpha}\z^0)$.

Let $\delta_{\alpha} Y_t=\Delta_{\alpha} \y_t-\partial_{\alpha}\y_t^0, 
\delta_{\alpha} Z_t=\Delta_{\alpha} \z_t-\partial_{\alpha}\z_t^0$, and
\begin{align*}
f_{1}(t, \Delta_{\alpha} \y_t)&=A_{\alpha}(t)\Delta_{\alpha} \y_t+\varphi_{\alpha}(t),\\
f_{2}(t, \partial_{\alpha}\y_t^0)&=\partial_{y}f(0, t,\y_{t}^{0})\partial_{\alpha}\y_t^0+\partial_{\alpha}f(0, t,\y_{t}^{0}).
\end{align*}
Applying It\^{o}'s formula to $|\delta_{\alpha} Y_t|^p$, after localization, we can get
\begin{align*}
&|\delta_{\alpha} Y_t|^p+\frac{p(p-1)}{2}\int_{t}^{T}|\delta_{\alpha} Y_s|^{p-2} |\delta_{\alpha} Z_s|^2 ds\\
= &|\frac{g(\xi^{\alpha})-g(\xi^{0})}{\alpha}-g'(\xi^{0})(\xi-\xi^0)|^p+p\int_{t}^{T}|\delta_{\alpha} Y_s|^{p-2} (\delta_{\alpha} Y_s) (f_{1}(s, \Delta_{\alpha} \y_{s})-f_{2}(s, \partial_{\alpha}\y_s^0))ds \\
&-p\int_{t}^{T}|\delta_{\alpha} Y_s|^{p-2} (\delta_{\alpha} Y_s) (\delta_{\alpha} Z_s)\cdot dB_s.
\end{align*}
Since $A_\alpha(s)\leq0$, then the drift term admits the decomposition
$$|\delta_{\alpha} Y_s|^{p-2} (\delta_{\alpha} Y_s) (f_{1}(s, \Delta_{\alpha} \y_{s})-f_{2}(s, \partial_{\alpha}\y_s^0))\leq |\delta_{\alpha} Y_s|^{p-1}|f_{1}(s, \Delta_{\alpha} \y_{s})-f_{2}(s, \partial_{\alpha}\y_s^0)|.$$
From the monotonicity condition, combined with Proposition 3.2 in Briand et al. \cite{briand2003lp}, it follows that
\begin{align}\label{eq4.9}
&\E_{\pp} \left[\underset{0\leq t \leq T}{\sup}|\delta_{\alpha} Y_t|^p+ \left(\int_{0}^{T}|\delta_{\alpha} Z_s|^{2} ds \right)^{\frac{p}{2}} \right] \nonumber\\
\leq &C_p \E_{\pp} \left[|\frac{g(\xi^{\alpha})-g(\xi^{0})}{\alpha}-g'(\xi^{0})(\xi-\xi^0)|^p\right] \nonumber \\
&+ C_p \E_{\pp} \left[\int_{0}^{T}|\delta_{\alpha} Y_s|^{p-1}| |f_{1}(s, \partial_{\alpha}\y_s^0)-f_{2}(s, \partial_{\alpha}\y_s^0)|ds \right].
\end{align}

By the convexity of $g$, it yields that 
$$0\leq\frac{g(\xi^{\alpha})-g(\xi^{0})}{\alpha}-g'(\xi^{0})(\xi-\xi^0) \leq g(\xi)-g(\xi^0)+K|g'(\xi^0)|.$$
The right-hand side belongs to $L^p(\pp)$, since $g\in(0,1]$ and $|g'(\xi^0)|\leq \gamma h'(\xi^0)$. 
Besides,
$$\lim_{\alpha \to 0} \frac{g(\xi^{\alpha})-g(\xi^{0})}{\alpha}=g'(\xi^{0})(\xi-\xi^0),$$
by the dominated convergence theorem, we obtain that
\begin{align}\label{eq4.10}
\lim_{\alpha \to 0} \E_{\pp} \left[|\frac{g(\xi^{\alpha})-g(\xi^{0})}{\alpha}-g'(\xi^{0})(\xi-\xi^0)|^p \right]=0.
\end{align}

Let $\Lambda_t=A_{\alpha}(t)-\partial_{y}f(0, t,\y_{t}^{0}),~t\in[0,T]$, again by Young's inequality, 
				\begin{align*}
					\begin{split}
						&\E_{\pp} \left[\int_{0}^{T}|\delta Y_s|^{p-1}| |f_{1}(s, \partial_{\alpha}\y_s^0)-f_{2}(s, \partial_{\alpha}\y_s^0)|ds \right] \\
						\leq  &c_p \E_{\pp} \left[\underset{0\leq t \leq T}{\sup}|\delta Y_t|^{p} \right] 
						+d_p \E_{\pp} \left[ \left(\int_{0}^{T}|\Lambda_s \partial_{\alpha}\y_s^0  +(\varphi_{\alpha}(s)-\partial_{\alpha}f(0, t,\y_{s}^{0}))| ds \right) ^p \right] \\
						\leq &c_p \E_{\pp} \left[\underset{0\leq t \leq T}{\sup}|\delta Y_t|^{p} \right]+2^{p-1}d_p \E_{\pp} \left[\int_{0}^{T}|\Lambda_s \partial_{\alpha}\y_s^0|^p ds \right] \\
						&+2^{p-1}d_p \E_{\pp} \left[\int_{0}^{T}|\varphi_{\alpha}(s)-\partial_{\alpha}f(0, t,\y_{s}^{0})|^p ds \right], 
				\end{split}	\end{align*}
				where $c_p$, $d_p$ are constants, and $c_p$ is chosen with $C_pc_p<1$. Here $C_p$ is the constant in \eqref{eq4.9}. 
				
Similarly, the concavity of $u$ implies
\begin{align*}
u(c_t)-u(c^0_t)+Ku'(c^0_t) 
\leq \frac{u(c^{\alpha}_t)- u(c^{0}_t)}{\alpha}-u'(c^{0}_t)(c_t-c^0_t) \leq 0.
\end{align*} The left-hand side belongs to $L^p(\pp)$. 
Since				
\begin{align*}
\lim_{\alpha \to 0} \varphi_{\alpha}(t)&=\lim_{\alpha \to 0} \frac{-\gamma (\y^{0}_t)^q[u(c^{\alpha}_t)- u(c^{0}_t)]}{\alpha}=-\gamma(\y^{0}_t)^{q}u'(c^{0}_t)(c_t-c^0_t)=\partial_{\alpha}f(0, t,\y_{t}^{0},\z_{t}^{0}),
\end{align*}
then, by the dominated convergence theorem,
\begin{align}\label{eq4.11}
\lim_{\alpha \to 0} \E_{\pp}\left[\int_{0}^{T}|\varphi_{\alpha}(s)-\partial_{\alpha}f(0, t,\y_{s}^{0})|^p ds \right]=0.
\end{align}

Now we focus on the item $|\Lambda_t \partial_{\alpha}\y_t^0|^p$. By mean value formula, one can get
\begin{align*}
A_{\alpha}(t)=\int_{0}^{1}\partial_{y}f(\alpha, t,\y_{t}^{0}+\lambda(\y^{\alpha}_t-\y^{0}_t))d \lambda =-\gamma q u(c^{\alpha}_t)\int_{0}^{1} \left(\y^0_t+\lambda(\y^{\alpha}_t-\y^0_t) \right)^{q-1}d \lambda.
\end{align*}
By directly calculations, one has
\begin{align*}
\Lambda_t=&A_{\alpha}(t)-\partial_{y}f(0,t,\y^0_t) \\ 
=&-\gamma q u(c^{\alpha}_t) \int_{0}^{1} \left[(\y^0_t+\lambda(\y^{\alpha}_t-\y^0_t))^{q-1}-(\y^0_t)^{q-1} \right] d\lambda \\
&-\gamma q (u(c^{\alpha}_t)-u(c^{0}_t))(\y^0_t)^{q-1}.
\end{align*}By the results in step 1, when $\alpha \to 0$, it implies $\Lambda\to0$ in $dt\times d\pp$ measure.

Furthermore, 
$$|\Lambda_t|\leq 3\gamma q(u(c_t^\alpha)+u(c_t^0))\leq 6\gamma q(u(c_t)+u(c_t^0)).$$
From assumption A3 and $\beta>2ap$, then the right-hand side of the above inequality belongs to $L^{2p}([0,T]\times\Omega)$. Dominated convergence theorem implies that $\mathbb{E}_{\pp}[\int_0^T|\Lambda_t|^{2p}dt]\rightarrow 0$ as $\alpha$ goes to zero.

Coming back to the BSDE for $(\partial_{\alpha} \y^{0}, \partial_{\alpha} \z^{0})$, by the standard prior estimate for the monotonic type of BSDE, one can get 
\begin{align*}
&\mathbb E_{\pp}\left[\sup_{t\le T}|\partial_\alpha\bar Y^0_t|^{2p}
+\big(\int_0^T|\partial_\alpha\bar Z^0_t|^2dt\big)^p\right]\\
\le &C_p\mathbb E_{\pp}\left[
\left(|g'(\xi^0)(\xi-\xi^0)|+\gamma\int_0^T|u'(c_t^0)(c_t-c_t^0)|ds\right)^{2p}\right]<+\infty.
\end{align*}
Finally, one can get
\begin{align}\label{eq4}
\lim_{\alpha \to 0} \E_{\pp} \left[\int_{0}^{T}|\Lambda_t|^p |\partial_{\alpha} \y^0_t|^p dt \right] =0.	
\end{align}
				
In summary, combining $\eqref{eq4.9}$, $\eqref{eq4.10}$, $\eqref{eq4.11}$, and $\eqref{eq4}$, it shows that $(\Delta_{\alpha} \y, \Delta_{\alpha} \z)$ converges to $(\partial_{\alpha}\y^0, \partial_{\alpha}\z^0)$ in $\mathcal{S}^p \times \mathcal{M}^p$, as $\alpha$ tends to $0$.     \hfill$\Box$

\section*{C. Proofs for Subsection \ref{subsec:5.1}}
\addcontentsline{toc}{section}{C. Proofs of Lemmas}
\renewcommand{\theequation}{C.\arabic{equation}}
\setcounter{equation}{0}
\renewcommand{\thelemma}{C.\arabic{lemma}}
\setcounter{lemma}{0}

{\textbf{Proof of Lemma \ref{lem2}}.}    Let $\{(c^n,\xi^n)\}_{n\ge 1}\subset \mathcal{A}$ converge to $(c,\xi)$. We may assume without loss of generality that 
$$
\xi^n \to \xi,\quad \mathrm{d} \mathbb{P}\text{-a.s.},\qquad c^n \to c,\quad \mathrm{d}t \times \mathrm{d}\mathbb{P}\text{-a.e.}.
$$By Fatou's lemma, one can get 
$$
\mathbb{E}_{\mathbb{P}}\left[\xi^\beta+\int_0^T c_t^\beta \mathrm{d}t\right]\le \liminf_{n\to\infty}\mathbb{E}_{\mathbb{P}}\left[(\xi^n)^\beta+\int_0^T (c_t^n)^\beta \mathrm{d}t\right]<+\infty.
$$
Hence $(\hat{c},\hat{\xi})\in\mathcal{C}^\beta\times L_+^\beta(\mathcal{F}_T)$.  By assumption A3 and $2ap<\beta$, similar one can get that 
$$\mathbb{E}_{\mathbb{P}}\left[h'(\xi)^{2p}+h(\xi)^{2p}+\int_0^T u(c_t)^{2p}+u'(c_t)^{2p}\mathrm{d}t\right]<+\infty.$$Hence, $(c,\xi)\in\mathcal{A}$.  

For $1<r<\beta$. By Vitali's convergence theorem, one can get that $c^n\to c$ and $\xi^n\to\xi$ in $L^r([0,T]\times \Omega)$ and $L^r(\pp)$ respectively. On the other hand,  one has 
\begin{align*}
    |X_0^{c^n,\xi^n}-X_0^{c,\xi}|\leq\mathbb E_\pp\left[\tilde D_T|\xi^n-\xi|+\int_0^T\tilde D_t |c_t^n-c_t|dt\right].
\end{align*}By H\"older's inequality, one can easily obtain that $X_0^{c^n,\xi^n}\rightarrow X_0^{c,\xi}$ since $\tilde{D}$ has all the moments. Thus, we complete the proof. \hfill$\Box$

{\textbf{Proof of Lemma \ref{lem3}}.}  Firstly, for each $x>0$, $v<0$,  the strict convexity of $J(x,\cdot,\cdot,v)$ on $\mathcal{A}$ follows from the strict convexity of  $(c,\xi)\rightarrow \bar{Y}_0^{c,\xi}$ and the affine structure of budget equation.  

For any $\Q\in\mathcal{Q}_f$ and $(\tilde c,\tilde \xi)\in\hat{\mathcal{A}}(x)$, recall
$$F(\Q; \tilde c, \tilde \xi):=\mathbb E_{\pp}\left[(D_T^\Q)^qh(\tilde \xi)
+\int_0^T(D_t^\Q)^qu(\tilde c_t)dt\right]
+\frac1\gamma H_q(\Q\mid\mathbb \pp).$$
By Theorem \ref{th1}, one has that
$Y_0^{\tilde c, \tilde \xi}=\inf_{\Q\in \mathcal{Q}_f}F(\Q; \tilde c,\tilde \xi).$

Let $\{(c^n,\xi^n)\}_{n\ge1}\subset\mathcal{A}$\ converge to $(c,\xi)$ in measure. By Lemma \ref{lem2}, we have $(c,\xi)\in\mathcal{A}$. Without loss of generality, we assume 
$$
\xi^n\to\xi\ ~d\mathbb{P}\text{-a.s.},\qquad c_t^n\to c_t ~~~dt \times d\mathbb{P}\text{-a.e.}
$$
For the above given $(c,\xi)$, let $\Q^*$ be determined by Theorem \ref{th1}, hence $Y_0^{c,\xi}=F(\Q^*;c,\xi)$.  From the proof procedure, we observe that $D^{\Q^*}$ is bounded by \eqref{eq:dqt}. Therefore,  for the $\Q^*$, one can easily get  that 
$$F(\Q^*; c^n, \xi^n)\rightarrow F(\Q^*; c, \xi).$$

Since $Y_0^{c^n,\xi^n}\leq F(\Q^*; c^n, \xi^n)$, it implies that 
$$\limsup_n Y_0^{c^n,\xi^n}\leq F(\Q^*; c, \xi)=Y_0^{c,\xi}.$$The continuous decreasing map $y\rightarrow \exp_q(-\gamma y)$ gives that  
$$\bar Y_0^{c,\xi}\leq \liminf_n \bar Y_0^{c^n,\xi^n}.$$

Furthermore, by the definition of $\bar{J}(x, c^n, \xi^n, v)=\bar{Y}_{0}^{c^n, \xi^n}+v(x-X_0^{c^n,\xi^n})$ for all $n\geq 1$, we can derive that 
$$\bar J(x,c,\xi,v)\le\liminf_n\bar J(x,c^n,\xi^n,v)$$
since $X_0^{c^n,\xi^n}\rightarrow X_0^{c,\xi}$ by Lemma \ref{lem2}. The proof is complete.
\hfill$\Box$ 
          
{\textbf{Proof of Theorem \ref{th exist}}.} Consider the product reflexive Banach space
$$\mathbb{X}:=L^\beta\big(\Omega\times[0,T],\text{Prog},\,dt\times d\mathbb P\big)\,\times\, L^\beta\big(\Omega,\mathcal F_T,\mathbb P\big),\quad \beta>1.$$
From Assumption A4, the sequence $\{(c^n,\xi^n)\}$ is bounded in $\mathbb X$. Since $\mathbb X$ is reflexive, we can extract a subsequence (still denoted by $(c^n,\xi^n)_{n\geq1}$) such that
$$
(c^n,\xi^n)\rightharpoonup (c^*,\xi^*) \quad \text{weakly in } \mathbb X.$$

Applying Mazur's lemma (see \cite{Yosida}, Theorem 2, page 120) to $(c^n,\xi^n)_{n\geq1}\subset\mathbb X$, we can construct
$$
\big(c_n',\xi_n'\big):=\sum_{k=n}^{N_n}\lambda_{n,k}\big(c^k,\xi^k\big),
$$
satisfying $\big(c_n',\xi_n'\big)\to (c^*,\xi^*) \quad \text{strongly in } \mathbb X.$ Because $\hat{\mathcal A}(x)$ is convex, each convex combination satisfies $(c_n',\xi_n')\in \hat{\mathcal A}(x)$. By Assumption A4, every $(c_n',\xi_n')$ belongs to $\operatorname{co}\{(c^n,\xi^n)\}$, so
$\sup_{n\ge1}\mathfrak M(c_n',\xi_n')<+\infty.$

Since the mapping $(c,\xi)\to \overline Y_0^{c,\xi}$ is convex on $\mathcal A$. Applying Jensen's inequality for convex functionals yields that 
$$\bar Y_0^{c_n',\xi_n'}
=\bar Y_0^{\sum_{k=n}^{N_n}\lambda_{n,k}(c^k,\xi^k)}
\;\le\; \sum_{k=n}^{N_n}\lambda_{n,k}\,\bar Y_0^{c^k,\xi^k}.
$$
By definition of the infimum $\bar V(x)$, we have $\bar V(x)\le \bar Y_0^{c_n',\xi_n'}$. Since $\bar Y_0^{c^k,\xi^k}\to \bar V(x)$ as $k\to\infty$, for arbitrary $\varepsilon>0$, then there exists $n_0>0$ such that
$$
\big|\bar Y_0^{c^k,\xi^k}-\bar V(x)\big|<\varepsilon,\quad \forall\,k\ge n_0.$$

Therefore, for $n>n_0$, all summation indices obey $k\ge n>n_0$, and it implies that 
$$
\sum_{k=n}^{N_n}\lambda_{n,k}\bar Y_0^{c^k,\xi^k}
\le \sum_{k=n}^{N_n}\lambda_{n,k}\big(\bar V(x)+\varepsilon\big)=\bar V(x)+\varepsilon.
$$
Combining the above two inequalities, one gets 
$$
\bar V(x)\le \bar Y_0^{c_n',\xi_n'}\le \bar V(x)+\varepsilon,
$$
and sending $n\to\infty$, then $\varepsilon\to 0$, we obtain $\bar Y_0^{c_n',\xi_n'}\to \bar V(x).$
Therefore $(c_n',\xi_n')_{n'>n_0}$ is also a minimizing sequence for $\bar V(x)$.

Strong convergence in $L^\beta$ implies convergence in measure, i.e., 
$$\xi_{n}'\to\xi^*\ ~d\mathbb{P}\text{-a.s.},\qquad c_{n}'\to c^* ~~~dt \times d\mathbb{P}\text{-a.e.}
$$By Lemma \ref{lem2}, one deduce that $(c^*,\xi^*)\in\hat{\mathcal A}(x)$ and $X_0^{c_{n}',\xi_{n}'}\to X_0^{c^*,\xi^*}$. By the lower semicontinuity property in Lemma \ref{lem3} and $(c^*,\xi^*)\in\hat{\mathcal A}(x)$, one gets that 
$$
\bar V(x)\leq \bar Y_0^{c^*,\xi^*}\le \liminf_{n\to\infty}\bar Y_0^{c_n',\xi_n'}=\bar{V}(x).
$$
Hence, $\bar Y_0^{c^*,\xi^*}= \bar V(x)$, which means 
 $(c^*,\xi^*)$ attains the infimum for problem \eqref{eq4.1}, or equivalently in \eqref{eq4.4}.
Strictly convexity of $J(x,\cdot,\cdot,v)$ yields uniqueness of the minimizer within $\hat{\mathcal A}(x)$.  

It remains to prove $X_0^{c^*,\xi^*}=x$. We argue by contradiction. Assume $b:=X_0^{c^*,\xi^*}<x$. Construct a new candidate strategy
$$
\tilde c_t:=c_t^*,\qquad \tilde\xi:=\xi^*+(x-b).
$$
Since $h'$ is decreasing, $h'(\tilde\xi)\le h'(\xi^*)$, which gives $\mathbb E_{\pp}\big[h'(\tilde\xi)^{2p}\big]<+\infty$. Consumption is unchanged, so $(\tilde c,\tilde\xi)\in\mathcal A$. By construction $X_0^{\tilde c,\tilde\xi}=x$, hence $(\tilde c,\tilde\xi)\in\hat{\mathcal A}(x)$.

The utility $h$ is strictly increasing. By the strictly comparison theorem for BSDE \eqref{eq4.3}, one gets that 
$\bar Y_0^{\tilde c,\tilde\xi}<\bar Y_0^{c^*,\xi^*}=\bar V(x).$
This contradicts the definition of $\bar V(x)$. 
This completes the proof of Theorem \ref{th exist}.
 \hfill$\Box$           
\end{document}